\documentclass{jpp}
\usepackage{graphicx}
\usepackage{aas_macros}
\usepackage{xcolor}

\usepackage{csquotes}
\usepackage{hyperref}
\usepackage{lscape}
\usepackage{subcaption}

\usepackage{wrapfig}
\usepackage{rotating}
\usepackage{epstopdf}
\usepackage{float}

\usepackage[utf8]{inputenc}
\usepackage[T1]{fontenc}
\usepackage{amsmath}
\DeclareMathAlphabet{\mathbfsf}{\encodingdefault}{\sfdefault}{bx}{sl}

\newcommand{\tens}[1]{\mathbfsf{#1}}

\shorttitle{Effects of Mesh Topology on MHD Solution Features}
\shortauthor{Brchnelova et al.}
\title{Effects of Mesh Topology on MHD Solution Features in Coronal Simulations}

\author{M. Brchnelova \aff{1}
  \corresp{\email{michaela.brchnelova@kuleuven.be}},
 \and F. Zhang\aff{1},
 \and P. Leitner\aff{1},\aff{2},
 \and B. Perri\aff{1},
 \and A. Lani\aff{1},\aff{3}
 \and S. Poedts\aff{1,}\aff{4}}

\affiliation{\aff{1}Centre for Mathematical Plasma-Astrophysics, Department of Mathematics, Catholic University Leuven,
Leuven, Celestijnenlaan 200B, 3001 Leuven, Belgium
\aff{3}Von Karman Institute For Fluid Dynamics, Waterloosesteenweg 72. B-1640, Sint-Genesius-Rode, Brussels, Belgium
\aff{4}Institute of Physics, University of Maria Curie-Sk{\l}odowska, Pl.\ M.\ Curie-Sk{\l}odowska 5, 20-031 Lublin, Poland
\aff{2} Institute of Physics, University of Graz Universit\"atsplatz 5  8010 Graz, Austria}

\begin{document}

\maketitle

\begin{abstract}

Magnetohydrodynamic (MHD) simulations of the solar corona have become more popular with the increased availability of computational power. Modern computational plasma codes, relying upon Computational Fluid Dynamics (CFD) methods, allow for resolving the coronal features using solar surface magnetograms as inputs. These computations are carried out in a full 3D domain and thus selection of the right mesh configuration is essential to save computational resources and enable/speed up convergence. In addition, it has been observed that for MHD simulations close to the hydrostatic equilibrium, spurious numerical artefacts might appear in the solution following the mesh structure, which makes the selection of the grid also a concern for accuracy. The purpose of this paper is to discuss and trade off two main mesh topologies when applied to global solar corona simulations using the unstructured ideal MHD solver from the COOLFluiD platform. The first topology is based on the geodesic polyhedron and the second on UV mapping. Focus will be placed on aspects such as mesh adaptability, resolution distribution, resulting spurious numerical fluxes and convergence performance. For this purpose, firstly a rotating dipole case is investigated, followed by two simulations using real magnetograms from the solar minima (1995) and solar maxima (1999). It is concluded that the most appropriate mesh topology for the simulation depends on several factors, such as the accuracy requirements, the presence of features near the polar regions and/or strong features in the flow field in general. If convergence is of concern and the simulation contains strong dynamics, then grids which are based on the geodesic polyhedron are recommended compared to more conventionally used UV-mapped meshes.
\end{abstract}

\section{Introduction}
\label{sec:introduction}

%A long time ago, in a galaxy far, far away, and maybe even in our own galaxy actually, many potential civilisations have gone extinct because of the activity of their own Suns going haywire. Well, we better not become next... Funny story, but a bit too dramatic ;)

Since the first observations of the solar wind by the probes Luna 2 and Mariner 2 \citep{Gringauz1962, Neugebauer1962}, the scientific community has been trying to model its origin in the solar corona. Analytical solutions were first derived in simple cases such as the hydrodynamic limit in 1D \citep{Parker1958} or the magnetic case in 1D \citep{Weber1967} and 2D \citep{Sakurai1985}. It was not until the end of the 20th century that computational capabilities became sufficient to handle magnetohydrodynamic (MHD) numerical simulations, coupling the Navier-Stokes and the Maxwell's equations for simple configurations \citep{Endler1971, Pneuman1971, Keppens1999}. It became then possible to use directly magnetograms derived from observations as a lower boundary condition to obtain data-driven numerical simulations \citep{Mikic1996b,Usmanov1996,Linker1999}. The most elaborate coronal models nowadays are able to incorporate complex physics such as a realistic transition region, Alfvén waves and thermal conduction \citep{vanderHolst2014,Pinto2017,Mikic2018,Reville2020,Chhiber2021}. However the model presented here is at early stage of development, and will thus rely on the polytropic assumption (quasi-isothermal heating of the corona) \citep{Leitner}. All of these codes use the Finite Volume (FV) technique, taking advantage of the conservative form of the MHD equations to ensure a proper conservation of mass (with sometimes additional assumptions such as Reynolds averaging, see \cite{McComb1990} and \cite{Usmanov2011}).

%We simulate the solar corona using the equations of magnetohydrodynamics (MHD), which couple the Navier-Stokes equations and the Maxwell's equations. Several numerical methods for solving these equations exist, in particular the Finite Volume (FV) technique, which is also employed in this paper, can be considered state-of-the-art. 

The solar surface (when the transition region is not included, so at the beginning of the corona) is usually modelled as a perfect conductor with a prescribed surface magnetic field taken from a solar magnetogram. The domain is extended sufficiently outwards in the direction normal to the solar surface to allow the flow to become supersonic. In our application, this interface is also set according to the requirements of coupling with the EUropean Heliospheric FORecasting Information Asset (EUHFORIA) code \citep{Poedts} which is establishing itself as a standard tool in Europe for space weather predictions \citep{Pomoell2018,Scolini2018,Scolini2019}. The domain that has to be simulated is therefore a sphere of radius of roughly $21R_\text{Sun}$ with a spherical cut-out in the middle with a radius of $R_\text{Sun}$.It aims at providing key physical quantities for the coupling with heliospheric model EUHFORIA (see an application for example in \cite{Samara2021}). \textcolor{black}{At this stage, we only focus on steady-state global corona simulations with fixed magnetograms as photospheric boundary conditions. In the future, this work will be extended for time-dependent problems, where the magnetograms will be updated roughly every hour of physical time.}

Providing a high quality mesh is essential both for the accuracy of the solution as well as the convergence process. \textcolor{black}{If the mesh has locally high skewness, the numerical fluxes (which are computed from linearly reconstructed states), especially near the boundary where the mesh is more distorted, might be affected. Similarly, mesh size affects the amount of numerical dissipation in the domain. Thus, if we have cells of a very high aspect ratio, the numerical dissipation in each direction is effectively different, leading to the lack of consistency and possibly the lack also accuracy}. In addition to accuracy, MHD coronal codes also have to converge in reasonable times (for space weather operations, so at maximum a couple of hours), leading to various strategies in meshing the domain: choice between Cartesian and spherical coordinates, stretched grids, Adaptative Mesh Refinement (AMR), unstructured meshes, Yin-Yang grids \citep{Kageyama2004,Shiota2014}, to name a few. There are indeed several methods in which such a domain can be discretised.

In numerical codes which employ structured meshes, use is typically made of what can be referred to as a UV mapping \citep{Cosker}, i.e. a sphere obtained by mapping a 2D rectangular grid onto the spherical surface, with "U" and "V" denoting the axes of the 2D texture. This results in a spherical surface mesh defined by a number of lines of latitude and longitude. For our purposes, the spherical surface mesh is extended radially outwards, where the surface layer at each radius is equivalent to the one beneath it, but scaled. In practice, the complete 3D domain with such a topology can be discretised using three parameters; the number of sections in the longitudinal, latitudinal and radial directions. 

In geodesy, cartography or graphical modelling however, other approaches are also used. The UV-mapped sphere results in a mesh which has much finer elements near the poles compared to the equatorial regions. It is oftentimes useful to have roughly same-sized elements everywhere on the spherical surface, for which then the Goldberg–Coxeter construction can be utilised \citep{GOLDBERG}. In its simplest form, the latter gives the Goldberg polyhedron and the geodesic polyhedron. The surface of the Goldberg polyhedron consists of a combination of hexagons and pentagons, while the geodesic polyhedron consists of triangles. Although these configurations are usually discussed exclusively in the context of spherical surface meshing, it is easy to extend these surface grids to fill the entire 3D domain, just like the UV-mapped sphere discussed above.

In order to choose which of these two mesh topologies is the most suitable for which type of a simulation, the factors to consider must first be formulated.

Firstly, the mesh resolution distribution is an important factor. In order to model the solar weather features having the highest significance for the Earth, it is important that mainly the regions near the equator are well resolved in MHD simulations. 

Secondly, another important factor is how the mesh affects the convergence. \textcolor{black}{It should be noted that in this paper, the convergence of the solution is assessed by evaluating the residual in the solution computed as:}

\begin{equation}
    \mathrm{res}(a) = \log\sqrt{\sum_i\left(a_i^t - a_i^{t+1}\right)^2},
\end{equation}

\textcolor{black}{in which $a$ is the physical quantity of interest and $i$ and $t$ the spatial and temporal indices. Once the residual reaches a certain level, generally below -3 to -4 for the pressure and density, the solution no longer changes visibly between subsequent iterations and the convergence of the solver is assumed.} The coronal simulations in 3D are computationally very heavy as they consist of millions of elements and typically run for hours up to days even on High Performance Computing (HPC) systems. A poorly designed grid might significantly increase the required computational resources and even prevent the simulations from converging in the first place. \textcolor{black}{This is due to the fact that while the continuous values of the solution are solved in cell centre values, the spatial derivatives are represented through Gauss’ theorem as fluxes across the cell faces. Obviously, larger cells will lead to larger numerical dissipation, generally easing convergence, but also less physical solutions due to the missing resolution. Secondly, while the finite volume method can in principle use all polyhedral cell shapes, the reconstruction of the fluxes which happens on the cell boundaries and the related accuracy depend on the cell shapes and how well the face normals are aligned with the direction of the propagating waves.}

Thirdly, it should also be considered that some coronal simulations might require a different resolution than others. This could be the case for example when strong but small surface magnetic structures are present. In this case, it is advantageous if the mesh configuration allows for easy adaptability of its resolution. 

Finally, there is also another aspect to consider when dealing with gravitationally stratified MHD media. The numerical code must hold the hydrostatic equilibrium with a very good accuracy in order not to introduce spurious fluxes in the solution, as these might affect the actual physical behaviour of the structures of interest. However, a highly accurate numerical approximation of such flows (close to the hydrostatic equilibrium) might be very challenging for FV (finite volume) schemes because they introduce a truncation error and do not necessarily exactly preserve:

\begin{equation}
    \nabla p = -\rho \nabla \phi \quad \text{with} \quad \nabla^2 \phi = 4 \pi G \rho,
\end{equation}

\noindent in which $\phi$ is the gravitational potential, $\rho$ is the density and $p$ the pressure. A very high grid resolution might then be needed, making the convergence of such simulations challenging as well as expensive. Examples of such effects have been reported by, for example, \citet{Fuchs}, \citet{Popov} and \citet{Krause}. Similar phenomena were also observed in the simulations presented further on in this paper, since they are close or at hydrostatic equilibrium in addition to having their pressure and density profiles spanning several orders of magnitude.

\citet{Kaeppeli19} proposed corrections to their numerical scheme such that these effects are to some extent mitigated, which is what they refer to as making the scheme \textit{well-balanced}. However, the implementation of such corrections might be both time-consuming or even not entirely possible for all numerical solvers. In addition, these corrections have not yet been developed for all numerical schemes. If that is the case, choosing the correct grid to be as uniform as possible is essential to prevent such spurious fluxes due to these numerical inaccuracies (also here referred to as \textit{mesh artefacts}). With the correct selection of the mesh topology, these effects can be mitigated altogether in the majority of the domain, even when using standard schemes that are not well balanced. 

To summarise, when working on coronal MHD simulations dominated by hydrostatic equilibrium, the factors to consider when choosing the mesh topology and design are:
\begin{itemize}
    \item resolution distribution;
    \item resolution adaptability;
    \item convergence performance;
    \item size of the mesh artefacts.
\end{itemize}
These factors, which (to the Authors' knowledge) have never been addressed in details in available literature about MHD simulations of global solar corona, will be analysed and will be the main focus in this paper which is organised as follows:
\begin{itemize}
    \item Section \ref{sec:methods} discusses the mesh geometries considered in the study in more detail and introduces the numerical code used to run the coronal MHD simulations;
    \item Section \ref{sec:results} presents the numerical results corresponding to different mesh topologies along with the convergence histories, timings and an evaluation of the strength of the mesh artefacts;
    \item Section \ref{sec:discussion} analyses these results further and formulates recommendations depending on the mesh application;
    \item Section \ref{sec:conclusions} provides a summary of the conclusions. 
\end{itemize}

\section{Methods}
\label{sec:methods}

Now that the basic problem has been introduced, the two mesh configurations of interest will be revisited in more detail. Subsection \ref{subsec:topologies} will elaborate on the two topologies. Subsection \ref{subsec:generation} will discuss how the selected configurations are transformed into the full domains and the final types of grids generated for the simulations in this paper. Finally, Subsection \ref{subsec:CF} presents a short overview of the MHD solver which has been used for all computations in this work.

\subsection{Mesh Topologies}
\label{subsec:topologies}

\begin{figure}
  \centering
  \includegraphics[scale=0.18]{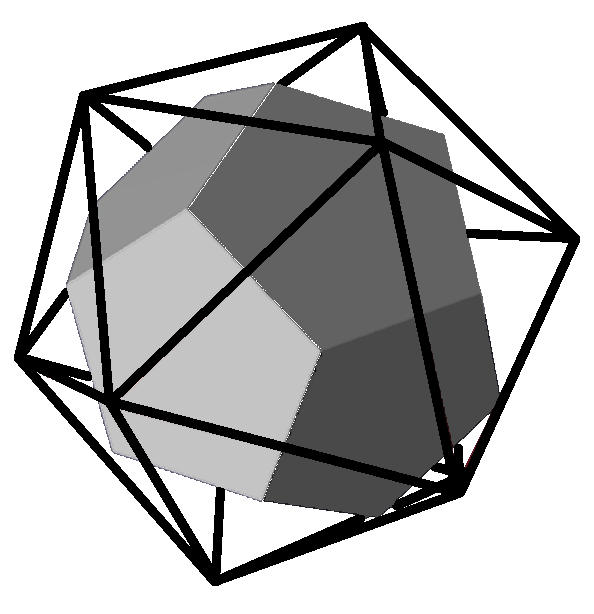}
  \caption{The comparison between the Goldberg polyhedron (solid) with hexagons and pentagons and the geodesic polyhedron (wireframe) with triangular elements. The Goldberg polyhedron is the dual of the geodesic polyhedron and vice versa.}
\label{fig:basisicosahedron}
\end{figure}

Two basic mesh topologies will be studied in this paper. The first topology can be retrieved from the Goldberg–Coxeter construction as the Goldberg or the geodesic polyhedron. Both the geodesic polyhedron and the Goldberg polyhedron are based on an icosahedron; whereas the Goldberg polyhedron is the dual of a Geodesic one and vice versa. This is illustrated in Figure \ref{fig:basisicosahedron}, where the Goldberg polyhedron is shown as a grey solid element and the geodesic polyhedron in a black wireframe. 

\begin{figure}
  \centering
  \includegraphics[scale=0.18]{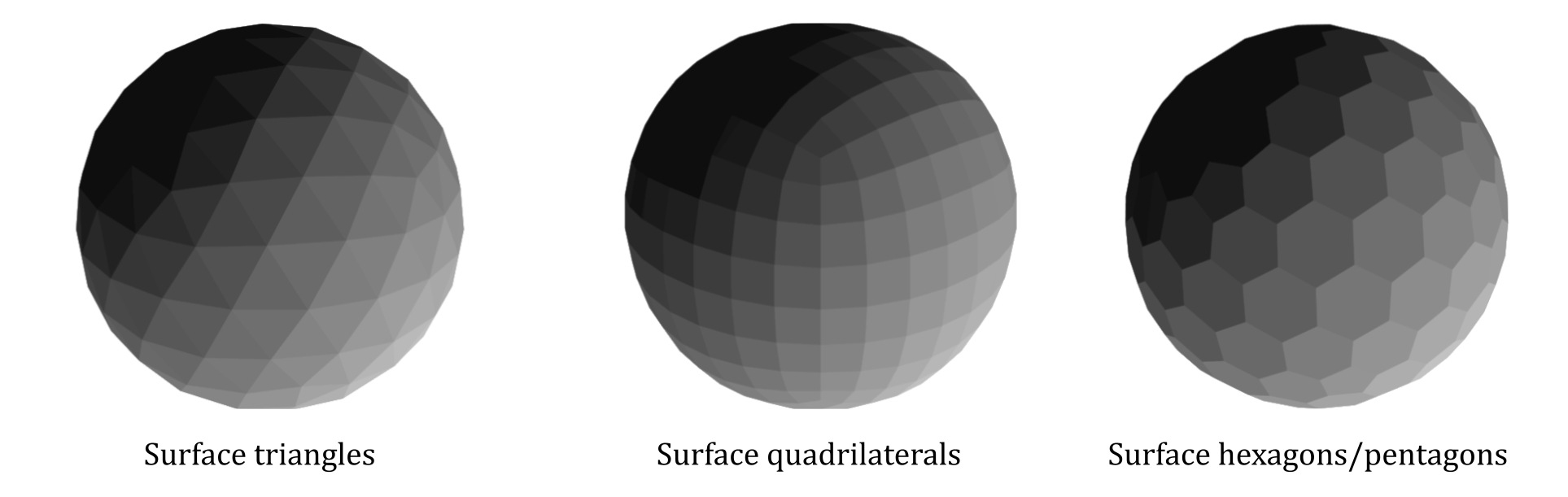}
  \caption{Three different configurations derived from the icosahedron-based spherical surface construction. In addition to the geodesic polyhedron and the Goldberg polyhedron, also an alternative with surface quadrilaterals is shown in the middle. The latter configuration, however, resulted in much stronger mesh artefacts in the solution compared to the geodesic polyhedron due to higher skewness, so it is not investigated further in this paper.}
\label{fig:icospheretypes}
\end{figure}

As seen from Figure \ref{fig:basisicosahedron}, the surface elements of the Goldberg and geodesic polyhedra are pentagons and hexagons in case of the former and triangles in case of the latter. An icosahedron-based spherical surface (from now on, referred to as an \textit{icosphere} for short) can also consist of quadrilateral elements. This can be preferable for some CFD numerical codes, since these quadrilateral surface elements in 3D result in \textcolor{black}{hexahedrals}. The three icospherical configurations discussed are shown in Figure \ref{fig:icospheretypes}.

\begin{figure}
  \centering
  \includegraphics[scale=0.18]{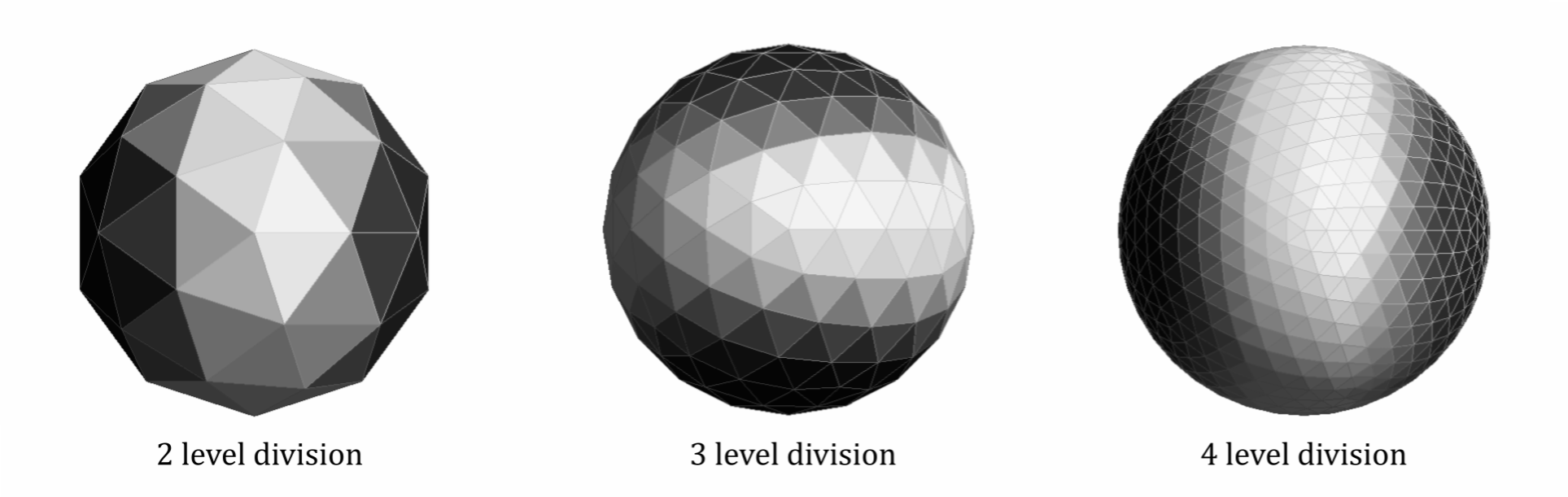}
  \caption{The principle of surface element splitting to achieve finer and better approximated spherical surface for the geodesic polyhedron. Each next level has four times as many elements on the surface as the one before. The division level for the grids used in this paper for the dipole and magnetogram simulations was 6.}
\label{fig:icosphere_division}
\end{figure}

Further refinement of the surface geometry is then possible via subdivision of the existing elements. An example of this for a triangle-based icosphere is shown in Figure \ref{fig:icosphere_division}, where the level-2 subdivision is the first subdivision of the basic construction from Figure \ref{fig:basisicosahedron}. This means that the surface resolution is limited to levels, where the next refinement level will have, in this case, four times as many elements as the previous level.

The second topology studied is the UV-mapped sphere, which is created from a 2D planar, regular mesh projected onto a spherical surface. This results in a spherical surface grid which has a certain number of lines of latitude and lines of longitude, usually with a constant angular spacing between them. While, by default, the majority of the surface elements are quadrilaterals, this projection is degenerate near the poles where the meridians meet, locally creating triangular elements (prismatic cells in 3D). 

\begin{figure}
  \centering
  \includegraphics[scale=0.25]{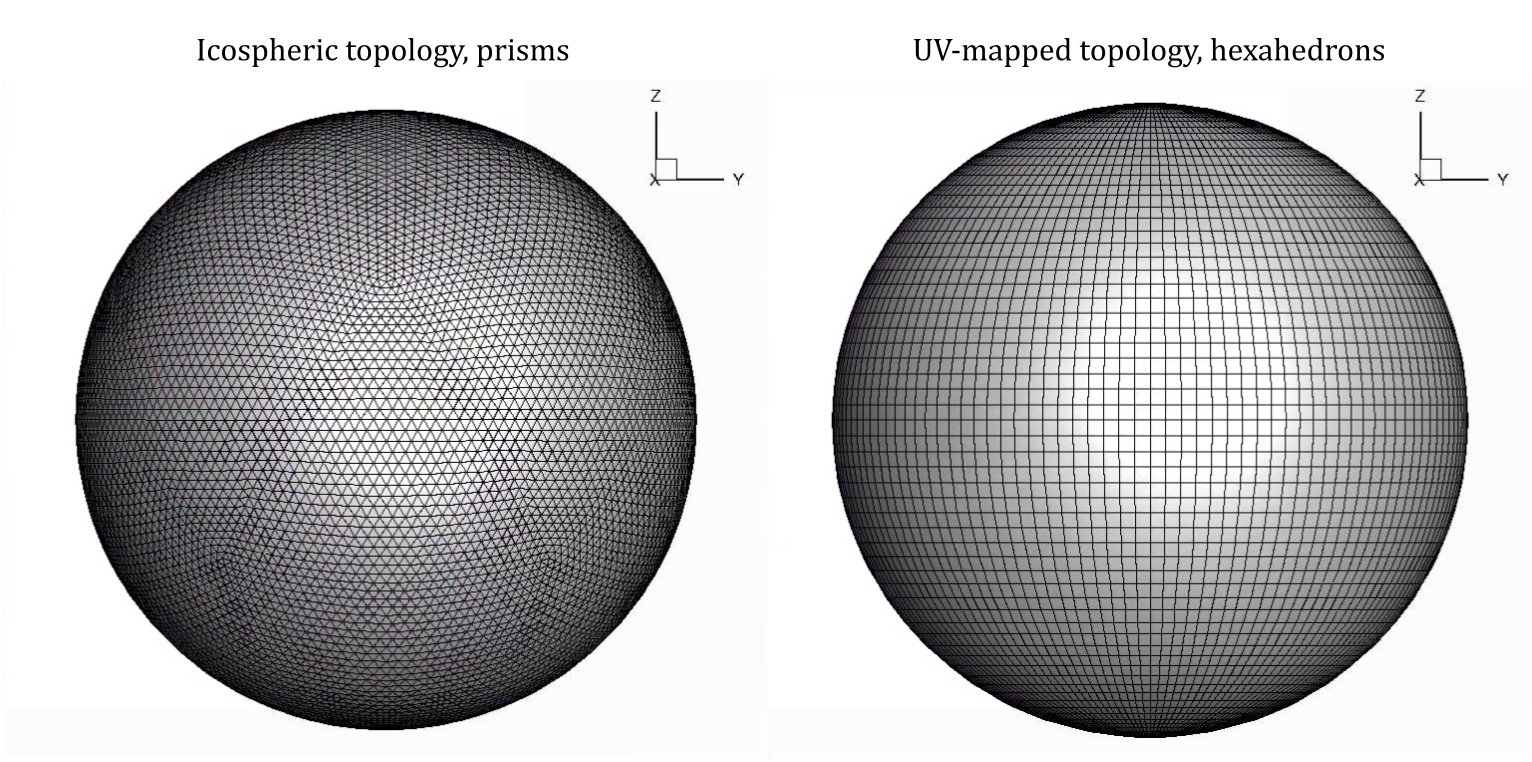}
  \caption{The two main surface topologies used to generate 3D meshes for the coronal MHD simulations further in the paper: the level-6 divided geodesic polyhedron (an \textit{icosphere} from here on) and the UV-mapped sphere (a \textit{UV sphere} from here on). }
\label{fig:topologies}
\end{figure}

The two topologies side by side are shown in Figure \ref{fig:topologies}. These are the surface grids that will be used in the rest of the paper. The reason why the geodesic polyhedron is picked instead of the Goldberg polyhedron is the fact that the CFD code used to run the simulations is not capable of handling heptahedrons (formed from pentagons) and octahedrons (formed from hexagons) which would be created should the Goldberg polyhedron be used for a construction of the full 3D domain. The quadrilateral-based icosphere is not further discussed because it is less uniform, so when compared to the triangle-based icosphere, the mesh artefacts were seen to be far more amplified. Since the CFD code used in present work can handle prism elements just as well as hexahedrons, there was no reason to not favour the triangle-based icosphere with weaker mesh artefacts.

\subsection{Mesh Generation}
\label{subsec:generation}

The surface meshes shown in the Figure \ref{fig:topologies} above were generated using Blender \footnote{{https://www.blender.org/}}. 
\textcolor{black}{Blender was chosen since it is powerful at visualisation even when it comes to large meshes, it is good at 2D mesh analysis, supported by all platforms and since the authors have past experience with this software package. Once generated, it was exported into the 
Stanford format, which is a polygon file format containing a simple description of the domain as a list of nominally flat polygons. The type of the polygons along with the list of boundary elements and normals are specified according to the format standard \footnote{{http://paulbourke.net/dataformats/ply/}}.}

\begin{figure}
  \centering
  \includegraphics[scale=0.30]{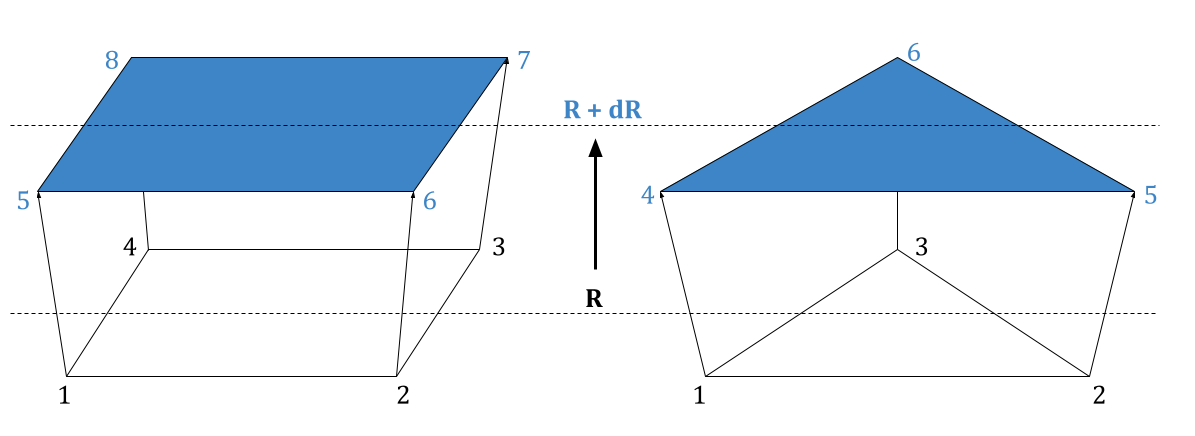}
  \caption{The principle of the extension of the spherical grids into a 3D domain applied in this work. The radial discretisation (from $R$ to $R + dR$) is independent of the selected topology. The 3D element is constructed using the original surface element (white), the extended surface element (blue) and the walls created by the radial extrusion.}
\label{fig:3dextension}
\end{figure}

Afterwards a python script was made to transform the surface geometries into a complete 3D domain according to a predefined radial discretisation function. The radial discretisation for these two grid types is independent of the topology. The principle of generating the 3D mesh from the surface mesh is shown in Figure \ref{fig:3dextension} for both triangular elements turning into prisms and quadrilateral elements turning into hexahedrons. The radius of the vertices of the basic element on the surface (white) is scaled according to $dR$ in the direction outward and the new surface element added. Then, the combination of this newly added element, the base element and the walls created by the radial extrusion are defined as the new stacked 3D cell. This stacking is applied for each new layer of surface elements, until the desired outer radius of $21 R_\text{Sun}$ is reached. Thus, the first and last layer of the 2D elements represent the inlet and outlet boundaries, respectively, for the computational domain. 

\begin{figure}
  \centering
  \includegraphics[scale=0.25]{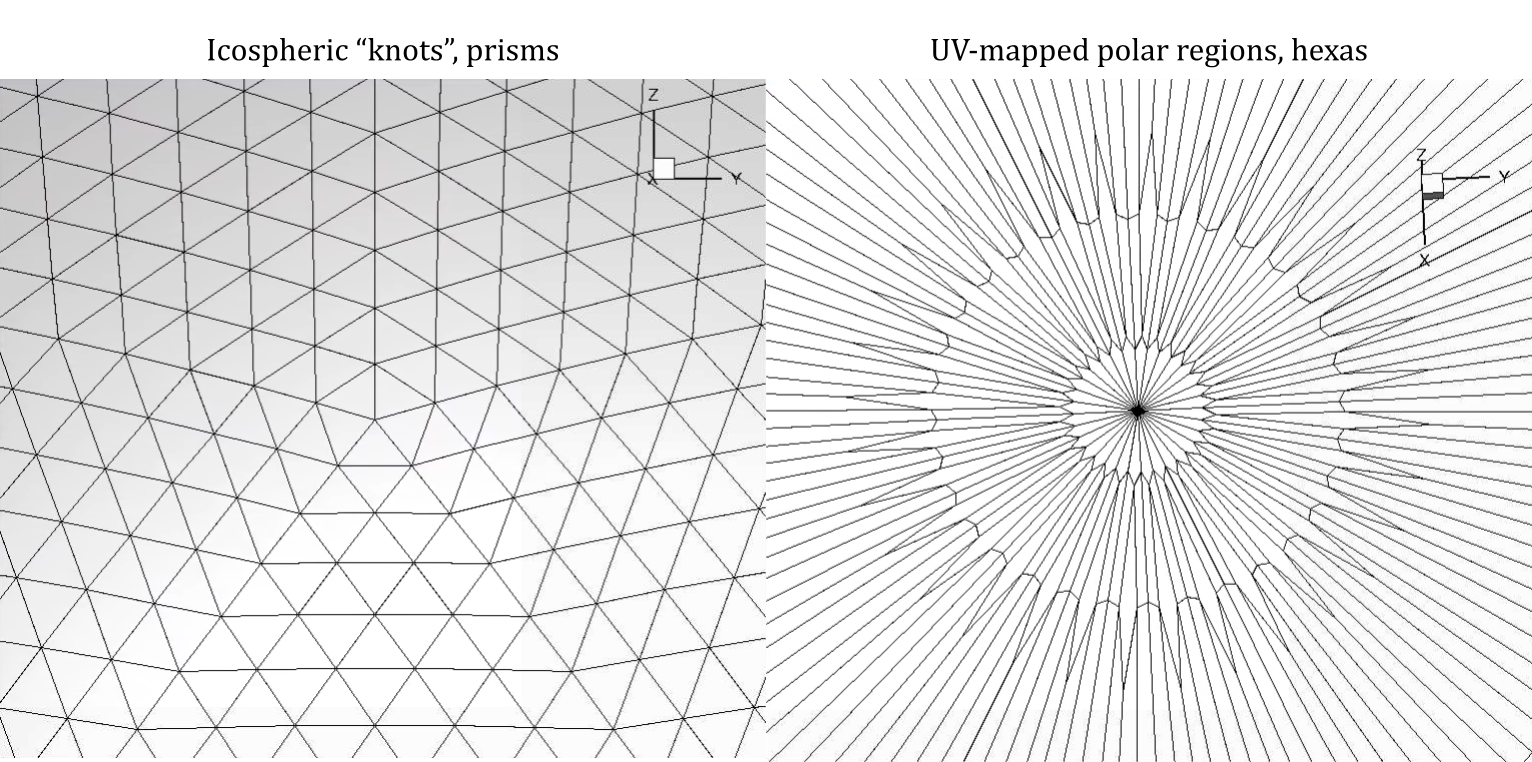}
  \caption{The non-uniformity regions in the two grid topologies. On the left, the \textit{knot} region in the icospheric grid is shown. This is the place where, in a Goldberg polyhedron, a pentagon would be located instead of a hexagon (and equivalently, here, the node only neighbours five other nodes instead of six).  On the right, the distorted polar regions of the UV mapped mesh are shown, where originally the degenerated prisms were placed.}
\label{fig:problems}
\end{figure}

The details about the grids used are in Table \ref{tab:meshes}. The basic UV mesh (\#1) was derived from the mesh commonly used by the Wind-Predict code \citep{Perri2020} for MHD simulations with real magnetograms \citep{Leitner}. This is a good reference of what resolution is generally required to sufficiently resolve coronal features and what type of UV sphere is used for it. The number of elements is 1.71M instead of 1.57M (192x64x128) because of the required handling of the polar regions to turn the degenerated prisms into hexahedrons, which results in additional elements being added on both sides of the domain, as shown later in Fig. \ref{fig:problems} (right). 

The second mesh is based on the geodesic polyhedron with a level-6 surface division. This division level was selected in order to have the surface element size close to the polar regions around the same size as for the UV mesh (\#1), which are the regions where these UV cells are the smallest. \textcolor{black}{This was done since the accuracy of the results is partly dependent on the numerical dissipation caused by the mesh, which is the smallest - and thus the most critical - for the most refined locations. Since especially later in the magnetogram test case we also focus on the flow behaviour in the polar regions, it is crucial to have sufficient (or at least comparable) minimum accuracy everywhere in the flowfield when comparing the grids.} However, since the icospheric mesh has the same mesh size almost everywhere whereas the default UV mesh has finer elements near the poles, this also means that the level-6 division mesh is much finer (roughly 3x) around the equator. Thus, when the same radial spacing is applied as in case of the UV mesh, the total number of elements is significantly larger (3.9M). 

Finally, having the same radial discretisation is important when comparing the structures in the velocity field between the solutions using the two mesh topologies. However, it would be bad practice to compare the convergence histories of these two grids (\#1 and \#2) in order to evaluate their performance, since higher overall numerical dissipation (here of the UV mesh, since it has locally much coarser elements than the icospheric one) makes the simulation easier to converge by default, regardless of the topology. Thus, another icospheric mesh (geodesic polyhedron) was created, also with a 6th level surface division, but with far fewer steps in the radial direction, to make comparison of convergence histories possible, with 1.3M elements (\#3).

\begin{table}
  \begin{center}
\def~{\hphantom{0}}
\begin{tabular}{llll}
    & Topology   & No. elements & Comments                                              \\
\#1 & UV         & 1.71M        & 192 sections radial, 64 latitudinal, 128 longitudinal \\
\#2 & Icospheric & 3.91M        & The same radial spacing as \#1                        \\
\#3 & Icospheric & 1.33M        & Similar cell volume near the equator as \#1          
\end{tabular}
  \caption{Overview of the grids used for the simulations.}
  \label{tab:meshes}
  \end{center}
\end{table}

\begin{figure}
  \centering
  \includegraphics[scale=0.25]{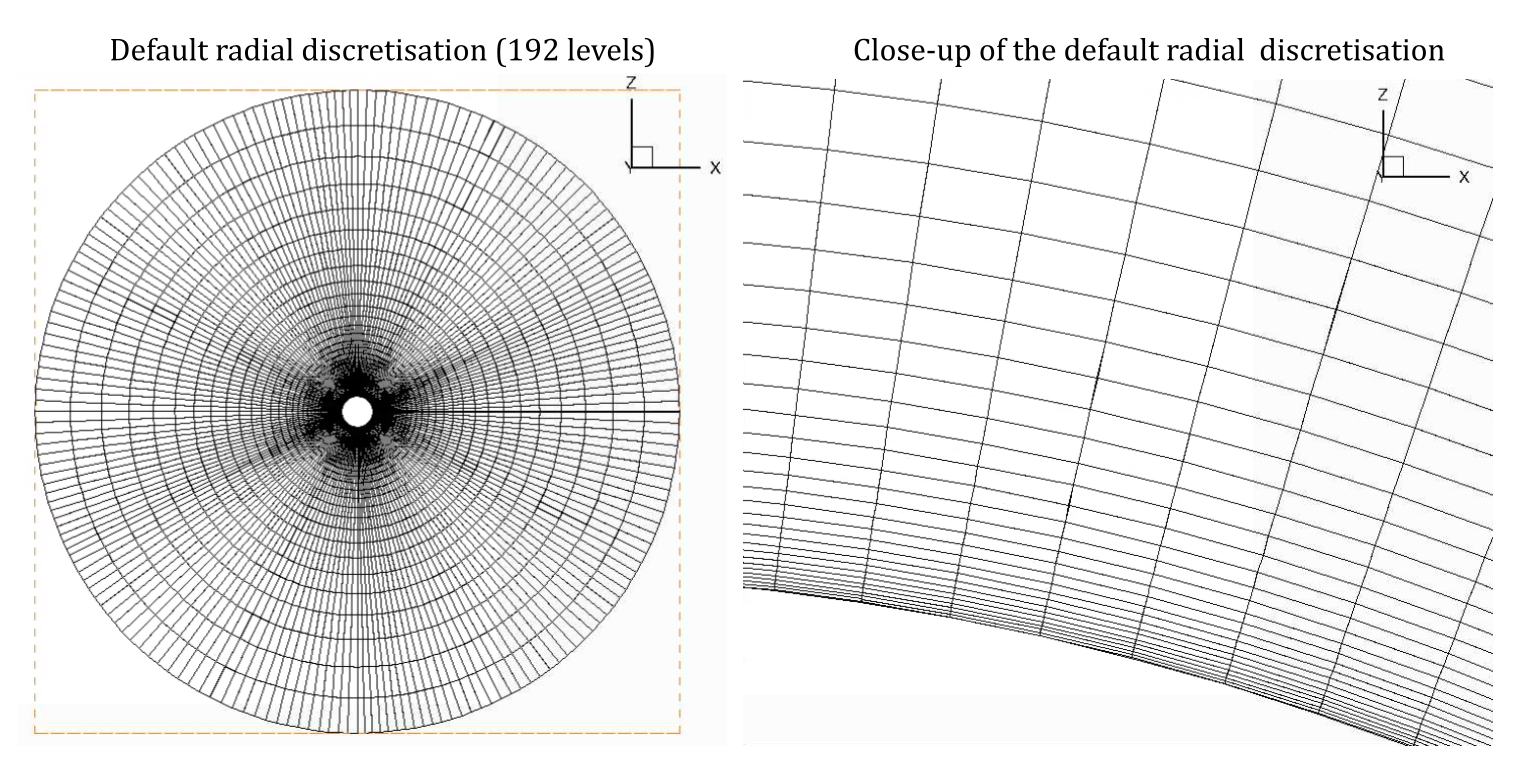}
  \caption{Illustration of the radial discretisation for the grids \#1 and \#2 and a close-up near the inner boundary.}
\label{fig:radial}
\end{figure}

The default radial discretisation for the first two grids along with a close-up near the inner boundary is shown in Figure \ref{fig:radial}. The cells are the finest near the inlet to resolve the magnetic field gradients properly.

\subsection{MHD Simulations}
\label{subsec:CF}

The COOLFluiD (Computational Object-Oriented Libraries for Fluid Dynamics) solver was used for all our CFD simulations. COOLFluiD is a framework for scientific HPC for multi-physics simulations \citep{Lani2005, Kimpe2005, Lani2013} with application to space re-entry flows \citep{Panesi2007, Degrez2009, Mena2015, Zhang2016}, radiation \citep{Santos2016}, magnetospheric/solar plasmas \citep{Laguna2016, Laguna2018, Asensio2019} and focus on algorithmic developments for high-speed flows \citep{Lani2011, Vandenhoeck2019}.   

Compared to state-of-the-art numerical tools for global coronal simulations, COOLFluiD-MHD \citep{Yalim, LaniGPU} is an \textcolor{black}{implicit} solver \textcolor{black}{(using a Backward Euler time discretization scheme)}, which means that Courant-Friedrichs-Lewy (CFL) numbers much higher than 1 (up to several thousands in some applications) can be afforded for converging to steady state. This makes the solution process much faster (up to 30x, see \cite{Leitner}), at the expense of increased memory requirements (which is not really an issue on modern HPC systems with hundreds or thousands of CPU-cores) when compared to time explicit solvers. In addition, COOLFluiD operates on unstructured grids, making it an ideal tool in order to study various mesh topologies and compare to structured UV grids. 

The code relies on a second-order accurate FV discretization for solving the ideal MHD equations with hyperbolic divergence cleaning in conservation form  and Cartesian coordinates (more details are given in \citet{Yalim, LaniGPU}):

\begin{equation}
\frac{\partial}{\partial t}\left(\begin{array}{c}
\rho \\
\rho \vec{v} \\
\vec{B} \\
E \\
\rho \\
\phi 
\end{array}\right)+\vec{\nabla} \cdot\left(\begin{array}{c}
\rho \vec{v} \\
\rho \vec{v} \vec{v}+\tens I\left(p+\frac{1}{2}|\vec{B}|^{2}\right)-\vec{B} \vec{B} \\
\vec{v} \vec{B}-\vec{B} \vec{v}+\underline{\tens I \phi} \\
\left(E+p+\frac{1}{2}|\vec{B}|^{2}\right) \vec{v}-\vec{B}(\vec{v} \cdot \vec{B}) \\
V_{r e f}^{2} \vec{B}
\end{array}\right)=\left(\begin{array}{c}
0 \\
\rho \vec{g}\\
0\\
0 \\
\rho \vec{g} \cdot \vec{v} \\
0
\end{array}\right),
\end{equation}

%\begin{equation}
%\frac{\partial \rho}{\partial t}+\nabla \cdot(\rho \boldsymbol{u})=0
%\end{equation}

%\begin{equation}
%\frac{\partial(\rho \boldsymbol{u})}{\partial t}+\nabla \cdot \boldsymbol{T}=\rho \boldsymbol{g}
%\end{equation}

%\begin{equation}
%\frac{1}{c} \frac{\partial \boldsymbol{B}}{\partial t}+\nabla \times\left(-\frac{\boldsymbol{u} \times \boldsymbol{B}}{c}\right)=\mathbf{0}
%\end{equation}

%\begin{equation}
%\frac{\partial}{\partial t}\left(\rho \frac{\boldsymbol{u}^{2}}{2}+\rho \mathcal{E}+\frac{\boldsymbol{B}^{2}}{8 \pi}\right) +\nabla \cdot\left[\left(\rho \frac{\boldsymbol{u}^{2}}{2}+\rho \mathcal{E}+P\right) \boldsymbol{u}+\boldsymbol{S}\right]=\rho \boldsymbol{g} \cdot \boldsymbol{u}
%\end{equation}

%with the stress tensor of

%\begin{equation}
%\boldsymbol{T}=\rho \boldsymbol{u} \otimes \boldsymbol{u}+\boldsymbol{I}\left(P+\frac{\boldsymbol{B}^{2}}{8 \pi}\right)-\frac{\boldsymbol{B} \otimes \boldsymbol{B}}{4 \pi}
%\end{equation}

%the Poynting vector by $\Vec{S} = -(\frac{1}{4\pi}) (\Vec{u} \times \Vec{B}) \times \Vec{B}$

in which \textcolor{black}{${E}$ is the total energy, $\vec{B}$ is the magnetic field, $\vec{v}$ the velocity, $\vec{g}$ the gravitational acceleration, $\rho$ the density,} and $p$ is the thermal gas pressure. The gravitational acceleration is given by $\vec{g}(r) = -(G M_\odot/r^2)\, \hat{\Vec{e}}_r$ and the identity dyadic $ \tens I = \hat{\Vec{e}}_x \otimes \hat{\Vec{e}}_x + \hat{\Vec{e}}_y \otimes \hat{\Vec{e}}_y + \hat{\Vec{e}}_z \otimes \hat{\Vec{e}}_z$. \textcolor{black}{All of the variables are non-dimensional. The magnetic field is adimensionalised by the value of $2.2\cdot 10^{-4}$ T ($B_{ref}$), the velocity field using the value of $ 4.8 \cdot 10^{5}$ m/s ($V_{ref}$), the density by $1.67\cdot 10^{-13}$ ($\rho_{ref}$), and the reference length is set to the Solar radius ($l_{ref}$).}To close the ideal MHD equations, the ideal equation of state is used. More information about the setup of the MHD solver including verification and validation can be found in \citet{Leitner}. The COOLFluiD-MHD solver is weakly coupled to COOLFluiD-Poisson, another FV code solving the Poisson equation on the same mesh in order to provide the Potential-Field Source-Surface (PFSS) corresponding to real magnetogram data ($B_r$) which are prescribed as inner boundary condition.  

The MHD boundary conditions are prescribed as follows. On the inner boundary, the values for $B_r$ and $B_\theta$ are prescribed according to the Potential-Field Source-Surface (PFSS) solution computed from the magnetogram values:

\begin{equation}
    B_{r,b,ND} =  \frac{{x_{b}}}{r_{b}} {B_{x,\text{PFSS}}} + \frac{{y_{b}}}{r_{b}}  {B_{y,\text{PFSS}}} + \frac{{z_{b}}}{r_{b}}  {B_{z,\text{PFSS}}},
\end{equation}

\begin{equation}
    B_{\theta,b} = \frac{{x_{b} z_{b}}}{\rho_{b} r_{b}} {B_{x,\text{PFSS}}}  + \frac{{y_{b} z_{b,ND}}}{\rho_{b} r_{b}} {B_{y,\text{PFSS}}} - \frac{\rho_{b}}{r_{b}} {B_{z,\text{PFSS}}},
\end{equation}

where the subscript $b$ indicates the boundary. The $\rho_{b}$ term is a geometric parameter given as $\sqrt{x_{b}^2 + y_{b}^2}$.

These spherical magnetic field components are then converted back to the Cartesian components $B_x, B_y$ and $B_z$ and using the inner state and the boundary state values defined above, the ghost cell values are computed.

For velocity, a small positive outflow is prescribed in terms of $V_r$ and $V_\theta$ in a way that any poloidal flux is removed:

\begin{equation}
    V_{r,b}^* = \frac{848.15}{(B_{ref}/\sqrt{\mu_0 \rho_{ref}})},
\end{equation}

\begin{equation}
    B_{\text{pol}} = \sqrt{B_{r,b}^2 + B_{\theta,b}^2},
\end{equation}

\begin{equation}
    V_{||} = V_{r,b}^* \frac{B_{r,b}}{B_{\text{pol}}^2}, \quad  V_{r,b} = V_{||} B_{r,b}, \quad V_{\theta,b} = V_{||} B_{\theta,b}.
\end{equation}

%\begin{equation}
%    V_{r,b,ND} = V_{||, ND} B_{r,b,ND}
%\end{equation}

%\begin{equation}
%    V_{\theta,b,ND} = V_{||, ND} B_{\theta,b,ND}
%\end{equation}

The $V_\phi$ component is set according to whether the simulation is with rotation on or not. By default, for a stationary simulation:

\begin{equation}
    V_{\phi,b} = 0.
\end{equation}

Just like in the case of the magnetic field, the transformation to the Cartesian coordinates then takes place and the respective $V_x, V_y$ and $V_z$ ghost cell values are prescribed.

The density and pressure on the boundary are set to $1.67\cdot 10^{-16}$ kg/m$^3$ and $4.16\cdot 10^{-2}$ Pa. 
The divergence cleaning term $\phi$ in the ghost cell is set such that its value is exactly 0 on the boundary:

\begin{equation}
     {\phi_{b}} = 0 \quad \xrightarrow[]{} \quad {\phi_{g}} = -{\phi_{i}},
\end{equation}

where the $g$ subscript refers to the ghost state and the $i$ subscript to the inner state. 

\begin{figure}
  \centering
  \includegraphics[scale=0.25]{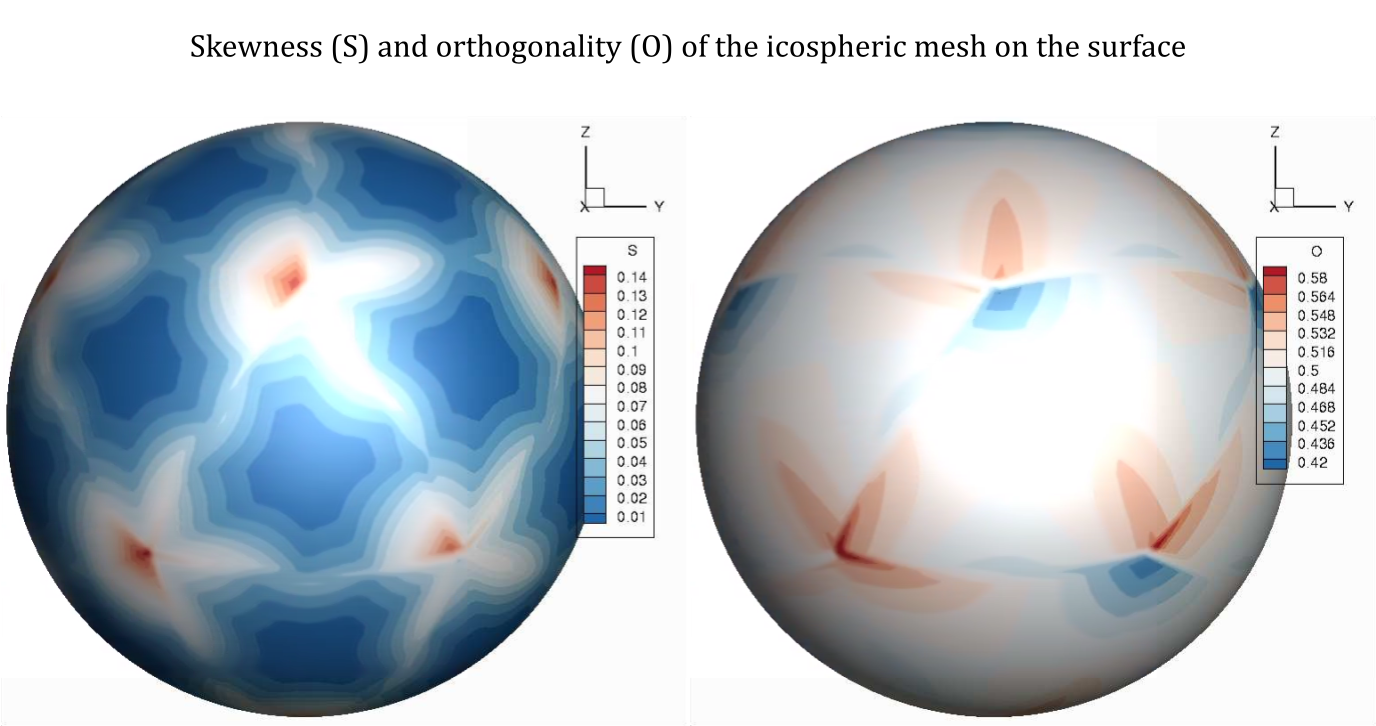}
  \caption{The skewness and orthogonality of the icospheric mesh, showing how these two quantities change around the mesh knots.}
\label{fig:SkewnessOrthogonality}
\end{figure}

On the outer boundary, the Neumann boundary conditions are prescribed. For the density and pressure, this is:

%\begin{equation}
%    {\rho_{g, ND}} = r_{i,ND}^2/r_{g,ND}^2 {\rho_{i,ND}}, \quad {p_{g, ND}} = r_{i,ND}^2/r_{g,ND}^2 {p_{i,ND}}
%\end{equation}

\begin{equation}
    {\rho_{g}} = {\rho_{i}}, \quad {p_{g}} = {p_{i}},
\end{equation}

%\begin{equation}
%    {p_{g, ND}} = r_{i,ND}^2/r_{g,ND}^2 {p_{i,ND}}
%\end{equation}

to ensure continuity of the temperature. The divergence cleaning constant $\phi$ is set to its reference value, typically 0:

\begin{equation}
    {\phi_{g}} = \phi_\text{ref}.
\end{equation}

The spherical components of the velocity field are assumed to be continuous:

\begin{equation}
    V_{r,g} = V_{r,i}, \quad V_{\theta,g} = V_{\theta,i}, \quad V_{\phi,g} = V_{\phi,i},
\end{equation}

%\begin{equation}
%    V_{\theta,g,ND} = V_{\theta,i,ND}
%\end{equation}

%\begin{equation}
%    V_{\phi,g,ND} = V_{\phi,i,ND}
%\end{equation}

which, just like in case of the inner boundary, is implemented by first the spherical-to-Cartesian and then Cartesian-to-spherical transformations. The azimuthal and polar components of the magnetic field are also assumed to be continuous and the radial component is scaled in the direction outwards:

\begin{equation}
    B_{r,g} = \frac{r_{i}^2}{r_{g}^2} {B_{r,i}}, \quad B_{\theta,g} = {B_{\theta,i}}, \quad B_{\phi,g} = {B_{\phi,i}}.
\end{equation}

%\begin{equation}
%    B_{\theta,g,ND} = {B_{\theta,i,ND}}
%\end{equation}

%\begin{equation}
%    B_{\phi,g,ND} = {B_{\phi,i,ND}}
%\end{equation}

\section{Results}
\label{sec:results}

The factors to consider when performing a mesh topology trade-off are, as introduced in Section \ref{sec:introduction}, the convergence performance, the resolution distribution, the adaptability and the ability of the mesh to minimise spurious numerical fluxes. This last aspect should be touched upon in more detail before proceeding onto the convergence of the MHD simulations and general performance, since the mesh non-uniformities giving rise to these spurious fluxes have not been discussed yet. 

None of the topologies is perfectly uniform. In case of the icosphere, most nodes have 6 neighbouring nodes connected to them apart from a few places (here referred to as \textit{knots}) where only 5 neighbouring nodes are present. These are the regions in the Goldberg polyhedron (the dual to the current geodesic configuration) where the elements are pentagons instead of hexagons. One such knot is visualised in Figure \ref{fig:problems} on the left side. The knot thus creates five lines, connecting it to other knots, on which the mesh lines change directions.

\begin{figure}
  \centering
  \includegraphics[scale=0.25]{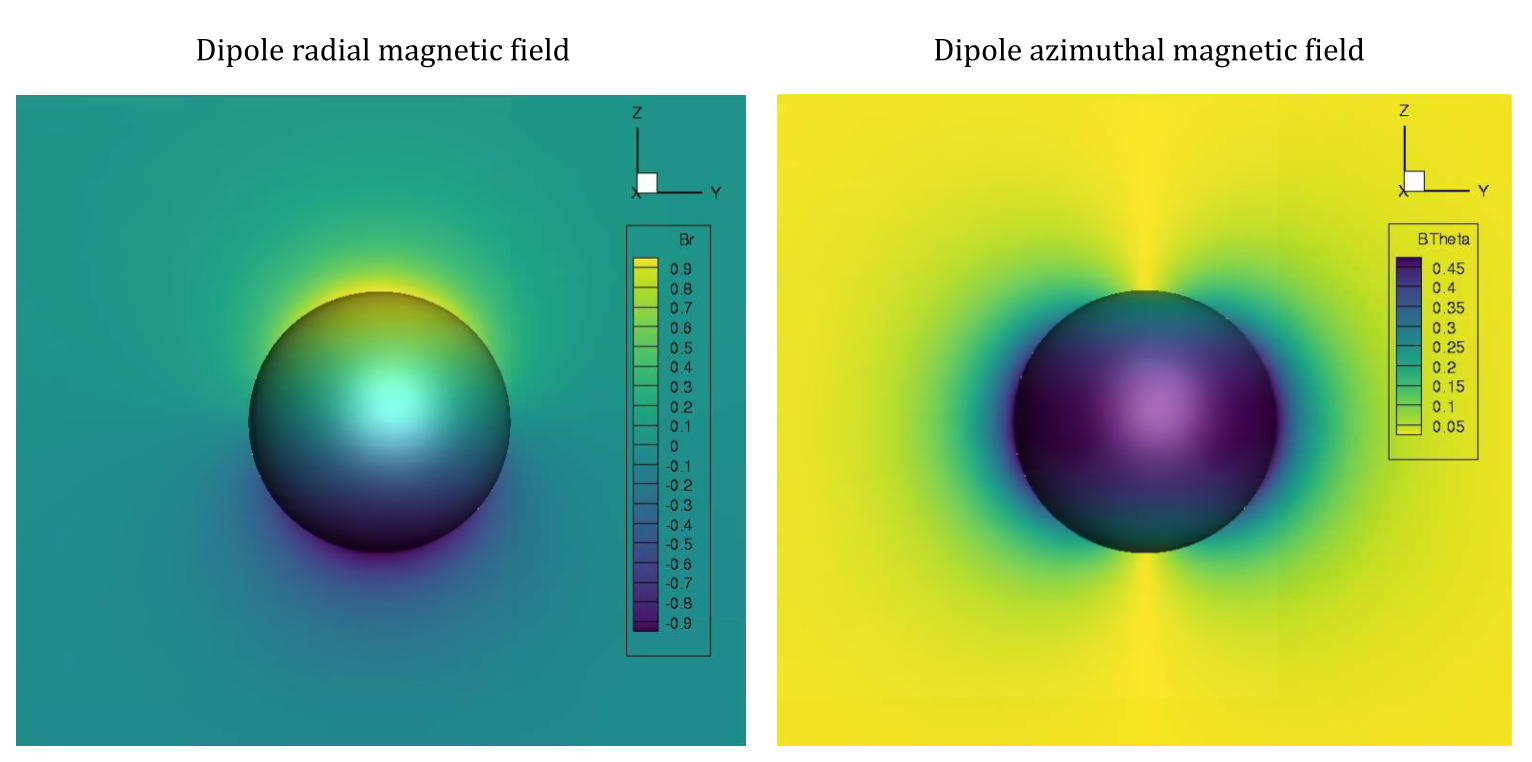}
  \caption{The configuration of the magnetic field to simulate the magnetic dipole, with the radial component $B_r$ on the left and azimuthal component $B_\theta$ on the right.}
\label{fig:dipoleconfigr}
\end{figure}

\textcolor{black}{The reason why these regions are problematic is the fact that mesh orthogonality and skewness here change a lot (in the longitudinal and latitudinal directions), see Figure \ref{fig:SkewnessOrthogonality}. With increased skewness, there is a bigger misalignment between the directions joining the centroids of the neighbouring cells and the face normals. This introduces excessive cross-dissipation and thus might change the solution. The solution is not necessarily better for a non-skewed mesh, since what matters for accuracy is whether the face normals are aligned with the propagating waves. However, this skewness can change the solution locally, which is what we see as these mesh artefacts. This increase in skewness is only in the latitudinal and longitudinal directions, not in radial. Thus, it is also expected that the errors will be observed mainly in the longitudinal and latitudinal components of the velocity, $V_\theta$ and $V_\phi$, around these knot regions.}

\begin{figure}
  \centering
  \includegraphics[scale=0.25]{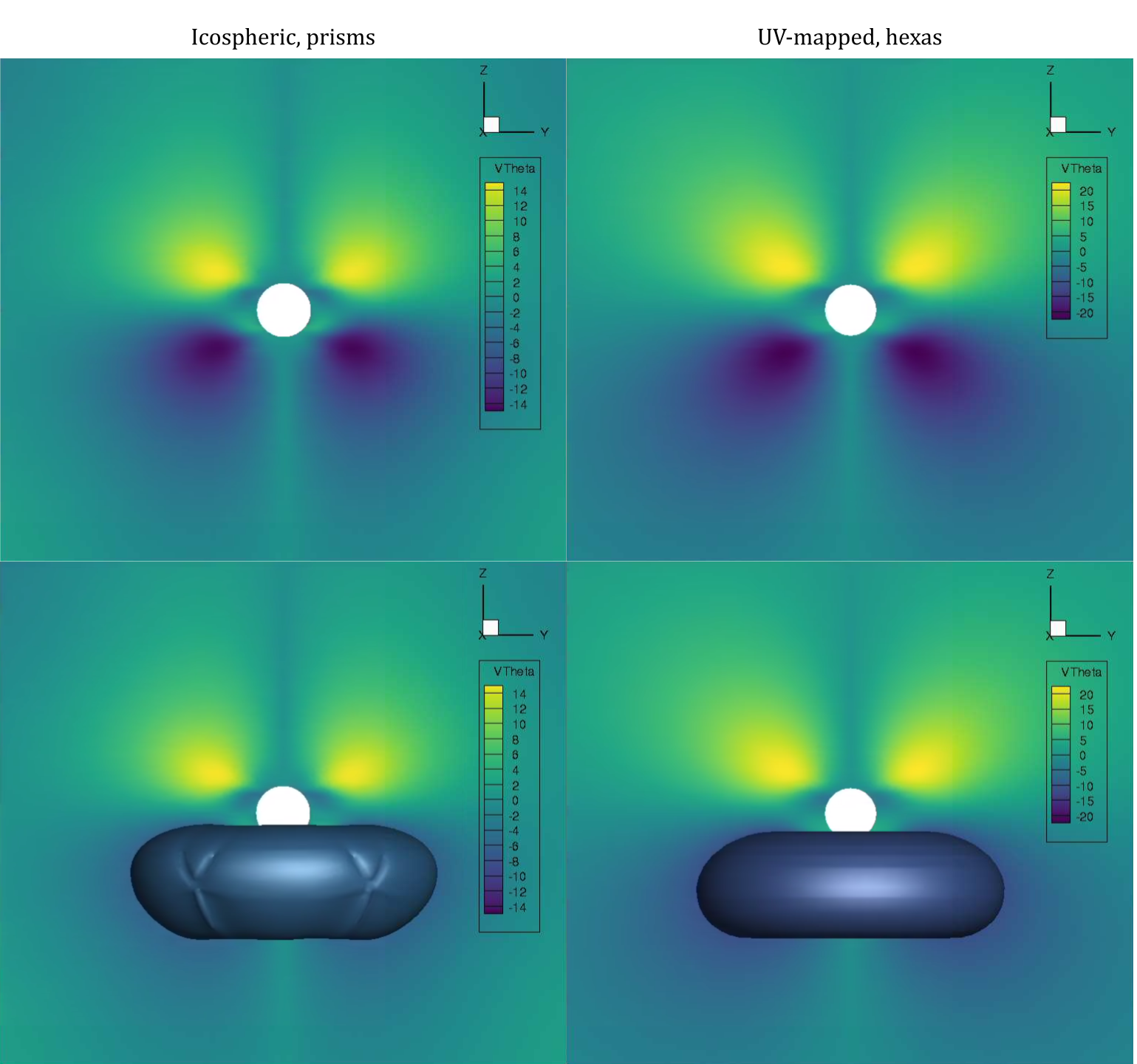}
  \caption{Comparison of the $V_\theta$ solution fields of a rotating dipole for the icospheric, prism based mesh (on the left) and the UV mesh with hexahedrons (on the right). While the solution projected onto the X axis looks similar, the isosurfaces of $V_\theta$ ($-9$km/s) show that the icospheric mesh produces mesh artefacts around the regions where the knots are present in the grid and where they are connected together.}
\label{fig:VthetaDipole}
\end{figure}

On the other hand, an non-adjusted UV sphere has degenerate elements near the poles, which means that in the full 3D domain, instead of hexahedrons, the polar regions are composed of prisms. In order to run simulations on fully hexaedral meshes instead of hybrid meshes (with both prismatic and hexahedral cells), the geometry of these prismatic polar regions was adjusted in a way to transform the prism cells into hexahedrons. This was achieved by adding new vertices and re-adjusting the existing cell edges. Several different techniques were attempted to achieve this result and, at the same time, to also keep the aspect ratio of the neighbouring cells at a reasonable value. The schematics of the configuration which was observed to produce the smallest mesh artefacts from the configurations tested is shown in Figure \ref{fig:problems} on the right. 

It is expected that the spurious fluxes will be mostly present in these non-uniform regions; around the knots in the icospheric mesh and around the polar regions in the UV mesh. 

\subsection{Dipole Artefacts}

Firstly, the simple case of a rotating magnetic dipole was computed since here, the possible mesh artefacts would be rather easy to identify. The magnetic field configuration and amplitude (in Gauss) is shown in Figure \ref{fig:dipoleconfigr}. The dipole was rotating with a prescribed boundary value of:

\begin{equation}
    V_{\phi,g} = 4.8 \cdot 10^{5} \Big(3.86 \cdot 10^{-3} r \sin(\theta) \Big).
\end{equation}

\begin{figure} 
  \centering
  \includegraphics[scale=0.25]{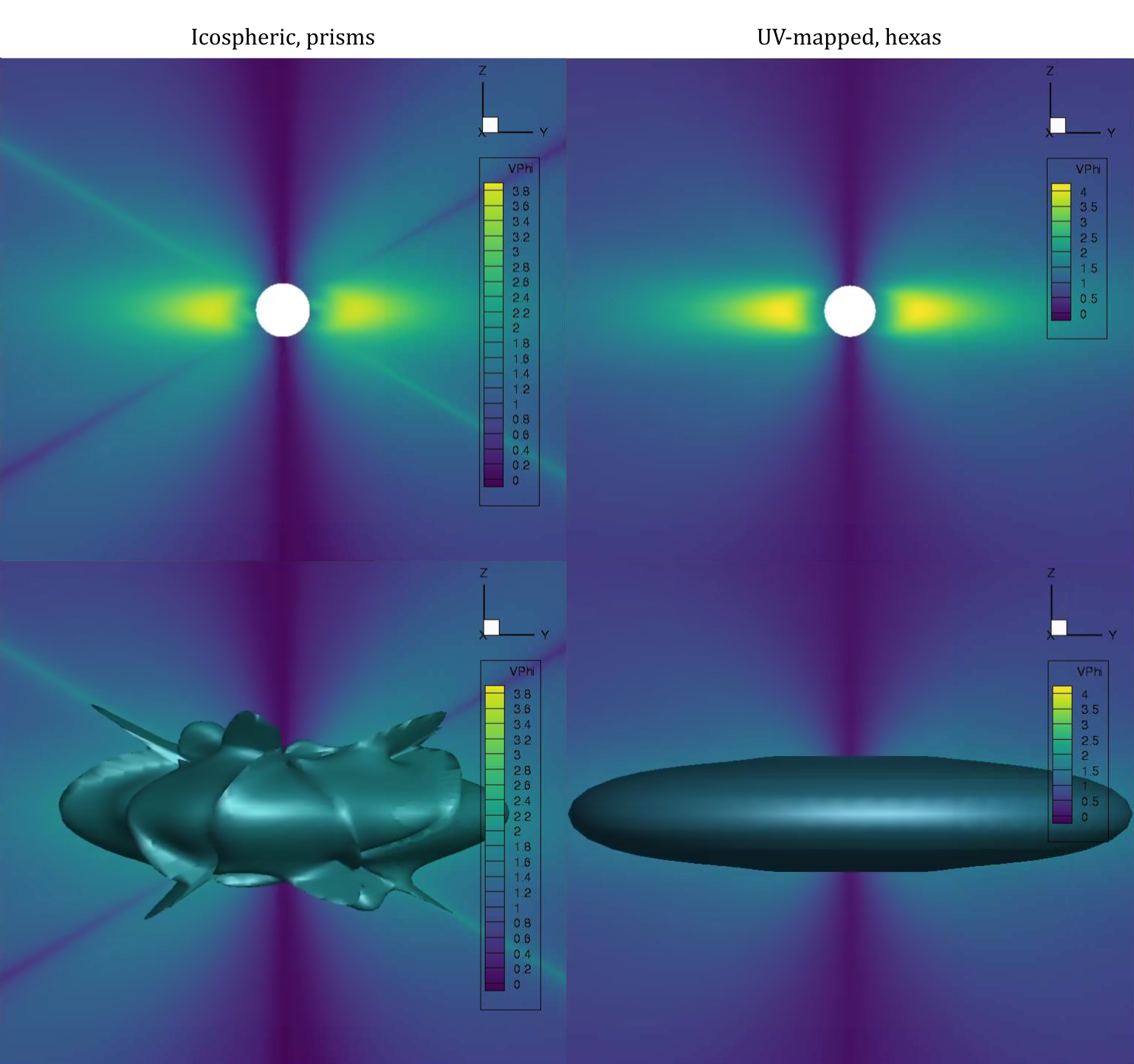}
  \caption{Comparison of the $V_\phi$ solution fields of a rotating dipole for the icospheric, prism based mesh (on the left) and the UV mesh with hexahedrons (on the right). In this case, the mesh artefacts from the knots and knot lines in the icospheric mesh are already observable in the projected field and the isosurface of $V_\phi$ (roughly $-1.5$km/s) is highly distorted.}
\label{fig:VphiDipole}
\end{figure}

The mesh artefacts were not observable in any of the solution fields apart from the azimuthal and polar velocity components, $V_\theta$ and $V_\phi$, as expected based on the above-introduced arguments. Indeed, here the artefacts could be observed in case of the icospheric mesh around the knot regions. Since the mesh of the UV sphere is completely uniform around the equatorial regions, where in this case the majority of the dynamics occurred, no artefacts were observable in the UV mesh results. The comparison between the two solutions for $V_\theta$ is shown in Figure \ref{fig:VthetaDipole}. While the results projected onto the X plane look the same, the mesh artefacts can be easily spotted when the isosurface of the $V_\theta$ component is plotted of roughly $-9$km/s. In this case, the location of the ripples in the isosurface corresponds exactly to the mesh knots.

A similar phenomenon can be observed, even more pronounced, in the polar velocity component, $V_\phi$. In this case, the icospheric mesh artefacts are even more amplified and can be identified even on the projected X plane, as displayed in Figure \ref{fig:VphiDipole}. The isosurface of $V_\phi$ of approximately $-1.5$km/s is in this case completely distorted. These artefacts did not diminish even with increased discretisation since even then, the mesh knots with high skeweness are still present.

\begin{figure}
  \centering
  \includegraphics[scale=0.25]{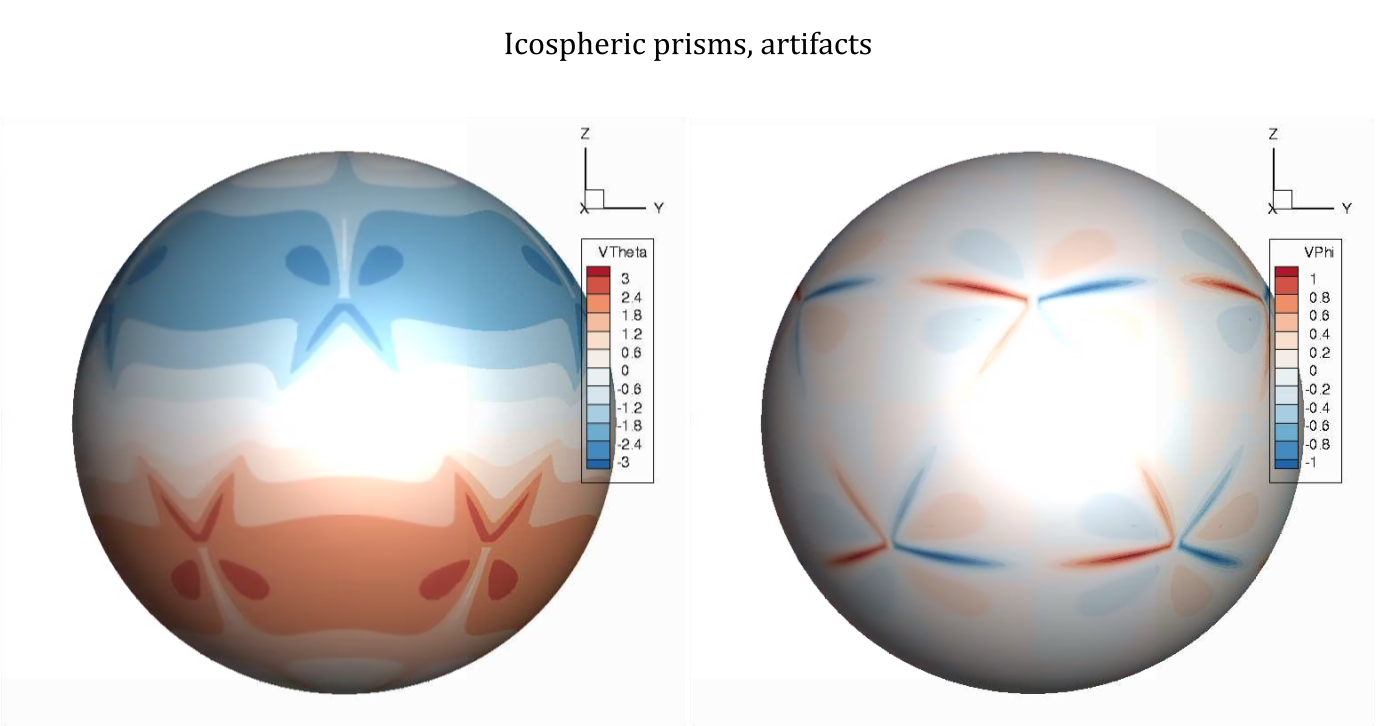}
  \caption{Projection of $V_\theta$ (left) and $V_\phi$ (right) icospheric solutions onto the outer boundary to better illustrate the mesh artefacts. The artefacts are contained in the lines connecting the knots of the mesh. }
\label{fig:outerboundary}
\end{figure}

A better visualisation of the icospheric artefacts can be made by projecting these two velocity components on the outer boundary, where they would otherwise have fairly smooth profiles. This is shown in Figure \ref{fig:outerboundary}. From this perspective, it is clearly noticeable how these regions of high distortion can be traced to the mesh knots and their connecting lines. 

From these plots on the outer boundary in Figure \ref{fig:outerboundary}, it was also determined that for the current configuration, locally, the worst case scenario errors in $V_\theta$ and $V_\phi$ reached 15-20\% and up to 30-40\%, respectively. Though these errors are contained in relatively small lines, they do significantly distort the local solution.

Before moving further however, it should be investigated how these spurious fluxes behave in simulations with stronger dynamics, which can be expected for more realistic magnetogram-based simulations. The rotating dipole was thus revisited, this time with double the imposed boundary value of $V_\phi$ to make it rotate faster and strengthen the existing physical velocity features. 

\begin{figure}
  \centering
  \includegraphics[scale=0.25]{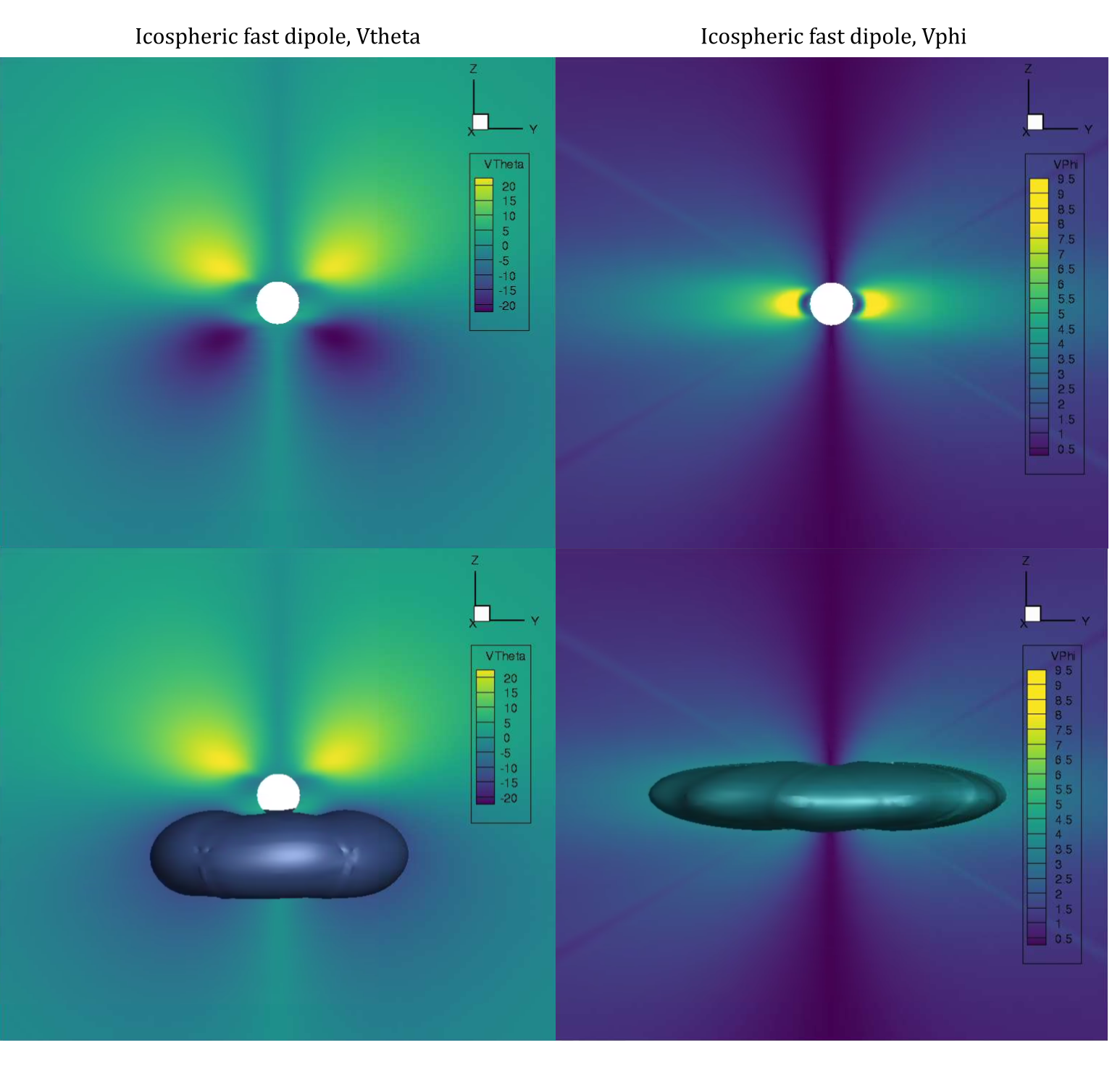}
  \caption{The $V_\theta$ and $V_\phi$ solution field X-plane projections and isosurfaces for a fast rotating dipole when the icopheric mesh is applied. It can be observed that when the dynamics in the solution is stronger (here the prescribed $V_\phi$ had double the magnitude compared to Figures \ref{fig:VphiDipole} and \ref{fig:VthetaDipole}), the mesh artefacts are weaker in the relative sense.}
\label{fig:DipoleFast}
\end{figure}

For the icospheric mesh, the results are shown in Figure \ref{fig:DipoleFast} in terms of $V_\theta$, $V_\phi$ and their respective isosurfaces. Especially in case of $V_\phi$ it is clear that the distortion is much less evident than it was in Figure \ref{fig:VphiDipole}, which is observable both on the X plane-projected field as well as on the smoothness of the isosurface. Indeed, from analysing these results, it is apparent that the effects of the spurious fluxes remain the same and do not scale with the actual, physical fluxes in the simulation. This means that for the MHD simulations with more pronounced solution features, such as those in which an actual surface magnetogram is used, the spurious fluxes due to the mesh knots could still be almost or completely negligible.  \textcolor{black}{For example, here, the relative error in $V_\phi$ decreased down to around 15\%. This is due to the fact that even with the increased $V_\phi$ from the faster rotation, the absolute errors in $V_\phi$ due to the artefacts remained roughly the same. This is also the case for most magnetogram-driven simulations.}

Thus, the magnetogram-based simulations were studied next.

\subsection{Magnetogram Artefacts}

\begin{figure}
  \centering
  \includegraphics[scale=0.25]{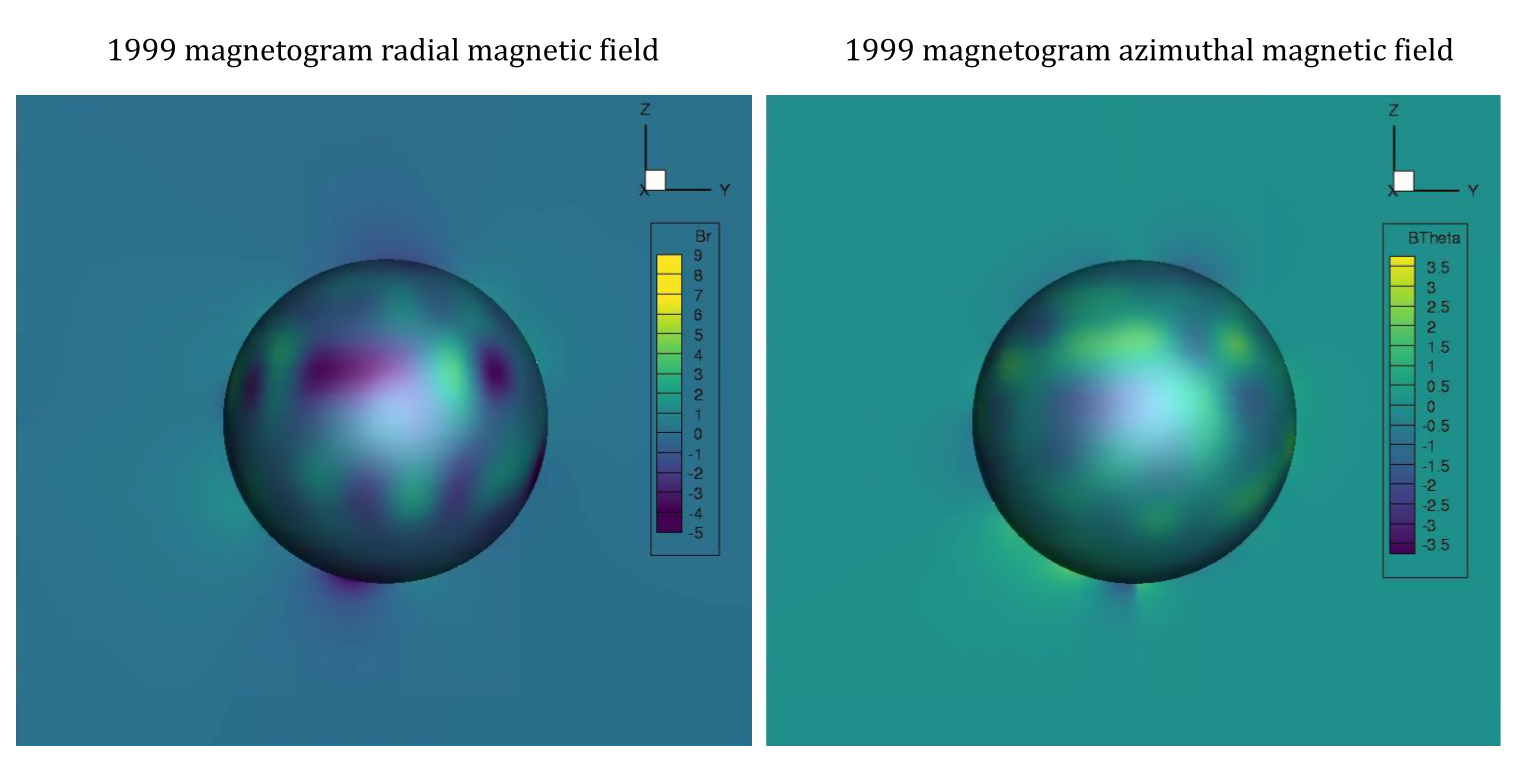}
  \caption{The radial and azimuthal component of the magnetic field, $B_r$ and $B_\theta$, applied on the inner boundary based on the solar magnetogram from 1999.}
\label{fig:MapConfig}
\end{figure}

To observe whether such strong spurious fluxes are also present in the results of more complex simulations with stronger dynamics, a data-driven test case was run. Here, the surface magnetic field was not dipole-like but a real solar magnetogram, from the solar eclipse of 1999 (near the solar maximum). The magnetic field configuration is shown in Figure \ref{fig:MapConfig}.

\begin{figure}
  \centering
  \includegraphics[scale=0.25]{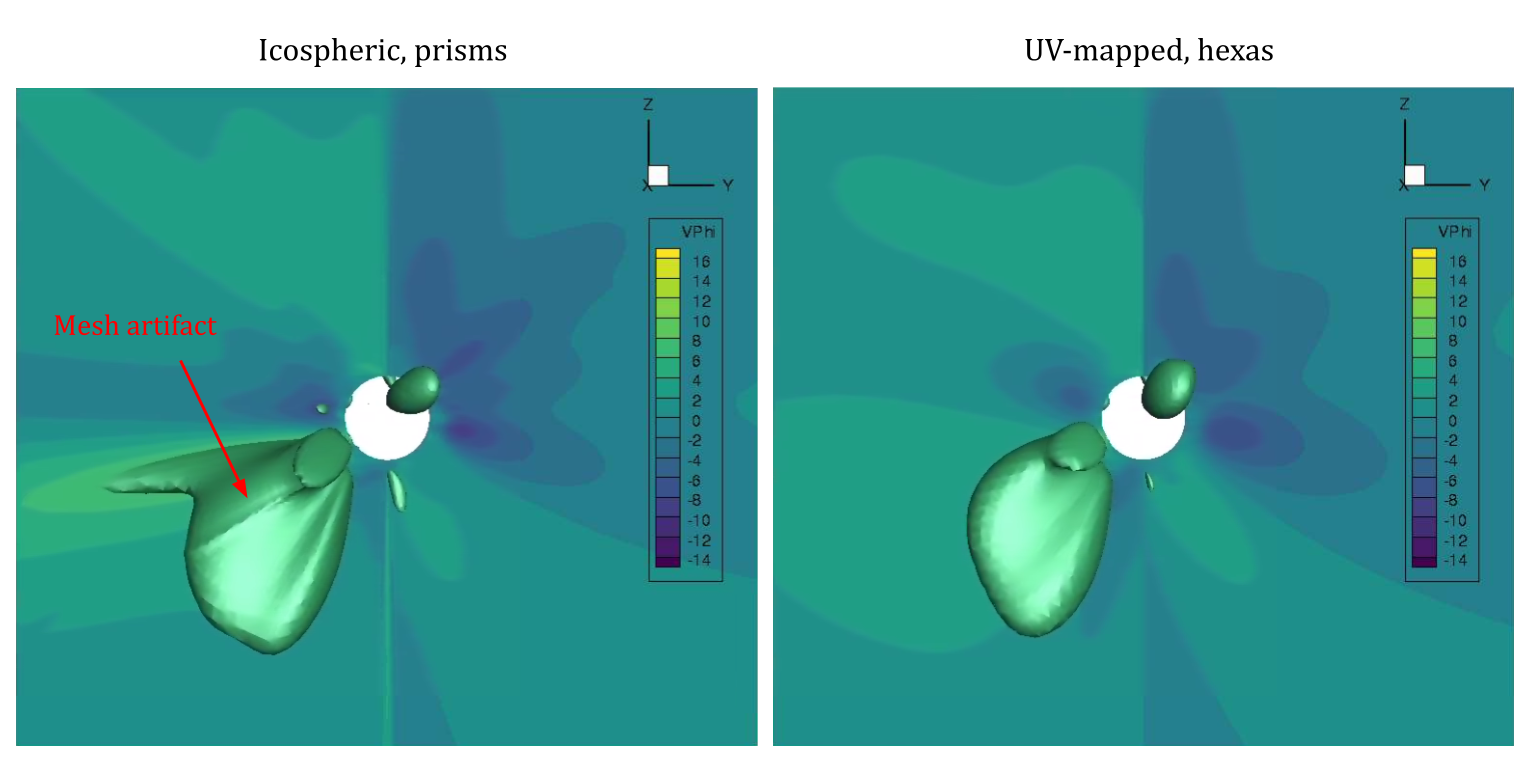}
  \caption{The 1.5km/s isosurface of $V_\phi$, showing the mesh lines in case of the icospheric grid (on the left) and a relatively clean solution of the UV grid (on the right). The two results are not completely equivalent since due to the different mesh topologies, the latitudinal and longitudinal resolution is significantly different.}
\label{fig:15Icosurface}
\end{figure}

\begin{figure}
  \centering
  \includegraphics[scale=0.25]{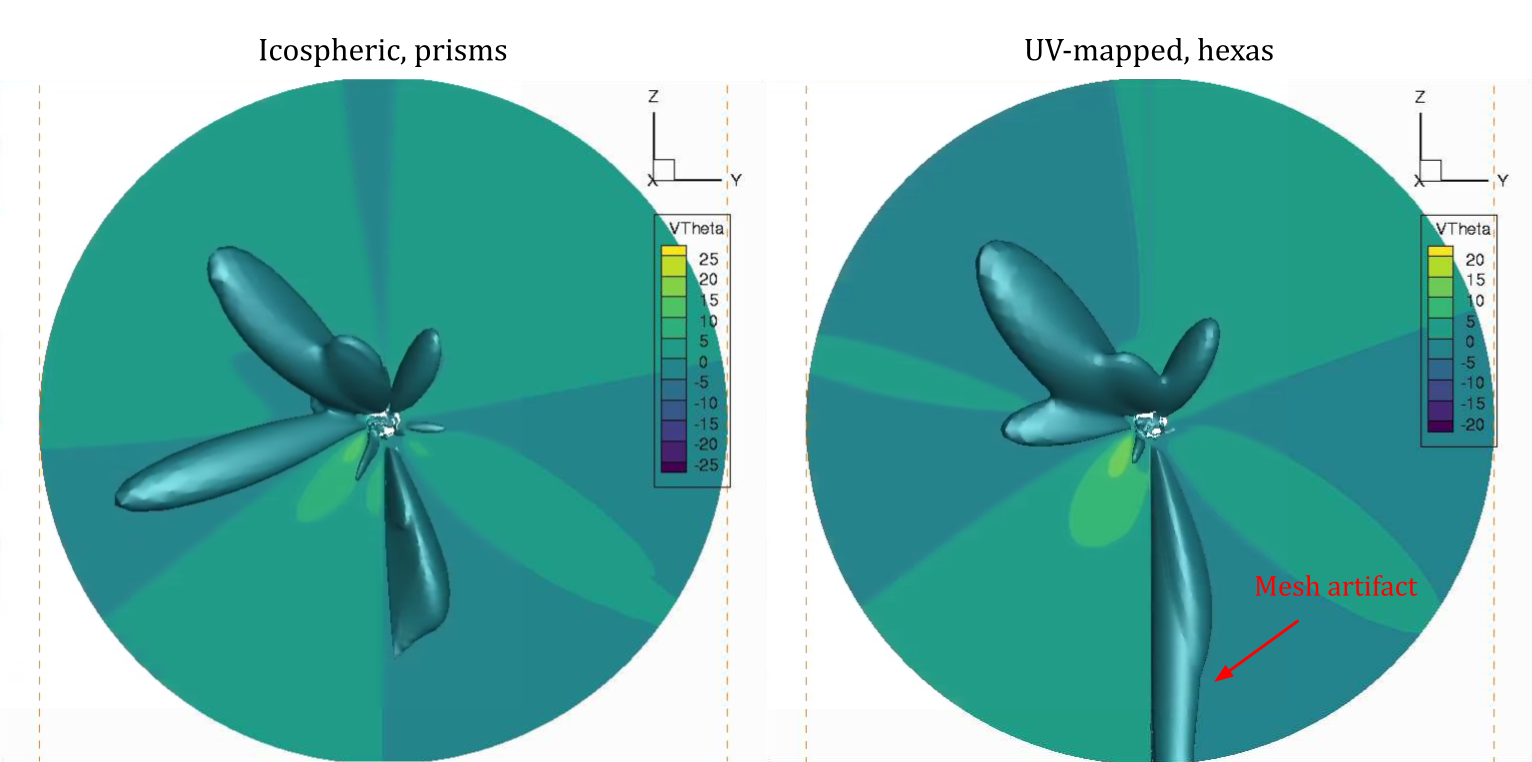}
  \caption{The -2.5km/s isosurface in $V_\theta$, showing that in this case, when the flow features are also present near the polar regions, the UV mesh with high distortion in these zones also creates very strong artefacts (on the right).}
\label{fig:25Icosurface}
\end{figure}

The $V_\phi$ and $V_\theta$ solution fields projected onto the X plane, including the isosurfaces, are shown in Figures \ref{fig:15Icosurface} and \ref{fig:25Icosurface}, respectively. Note that the contours are not exactly the same, even outside of the mesh-compromised regions. This is due to the fact that as mentioned in Section \ref{sec:methods}, the two topologies have a completely different surface resolution distributions, which affect the treatment of the surface magnetic fields and thus also the shape of the formed structures.

In this case, with complex flow dynamics, it is more difficult to determine which features are caused by the mesh non-uniformities. The easiest method is to directly compare the two results, since the topologies have the non-uniformities at different locations. When comparing the $V_\phi = -1.5$km/s isosurface in Figure \ref{fig:15Icosurface}, a clear knot line can be seen in the solution from the icospheric mesh, though its magnitude compared to the other features is much smaller than what was seen for the $V_\phi$ component of the rotating dipole. Here, this artefact barely compromises the accuracy of the simulation result.

Similar lines were not observed in the $V_\theta$ contours, which are shown in Figure \ref{fig:25Icosurface}. On the contrary, since in this case, the flow features were present everywhere including around the polar regions, the mesh artefacts were far more pronounced around the poles in case of the UV mesh where these regions are highly distorted.

\begin{figure}
  \centering
  \includegraphics[scale=0.25]{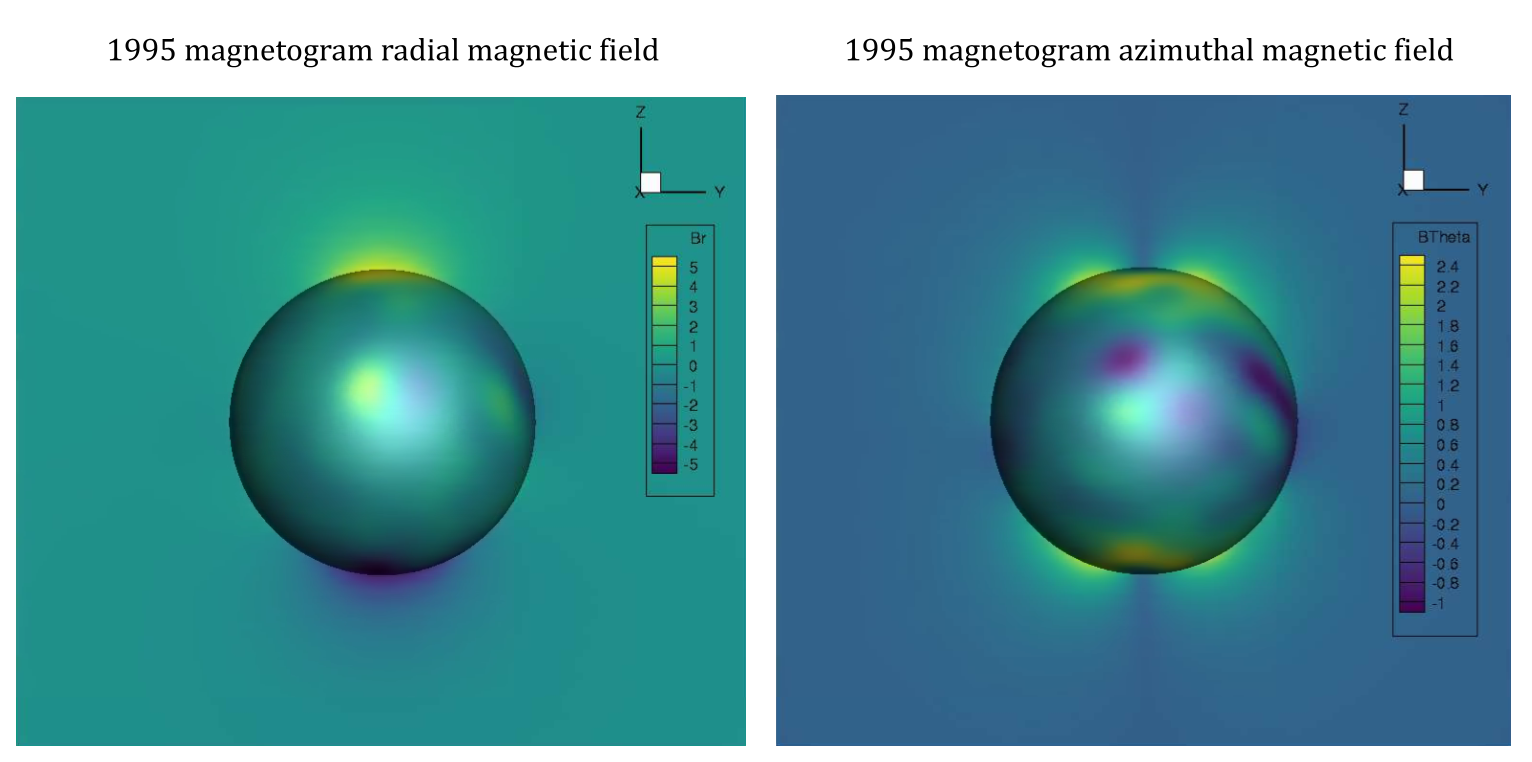}
  \caption{The radial and azimuthal component of the magnetic field, $B_r$ and $B_\theta$, applied on the inner boundary based on the solar magnetogram from 1995.}
\label{fig:1995Config}
\end{figure}

%Having now a good indication of the real appearance of the spurious fluxes in the solutions computed with the different mesh topologies, their numerical performance is discussed next.

\subsection{Numerical performance}

Having now a good indication of the real appearance of the spurious fluxes in the solutions computed with the different mesh topologies, their numerical performance is discussed next. In case of magnetogram-based simulations, the flow phenomena oftentimes also occur around the polar regions. In these regions, the distortion in the UV mesh is very high and therefore the convergence of the UV mesh is generally worse than that of the icospheric mesh. 

To illustrate this point, a magnetogram from 1995 was picked, the magnetic field configuration of which is shown in Figure \ref{fig:1995Config}. This magnetogram was selected for this purpose because it is from the solar minimum, when the magnetic field is similar to that of a simple dipole, and so there should be only very weak structures in the polar regions (and so the artefacts could be easily spotted).

\begin{figure}
  \centering
  \includegraphics[scale=0.25]{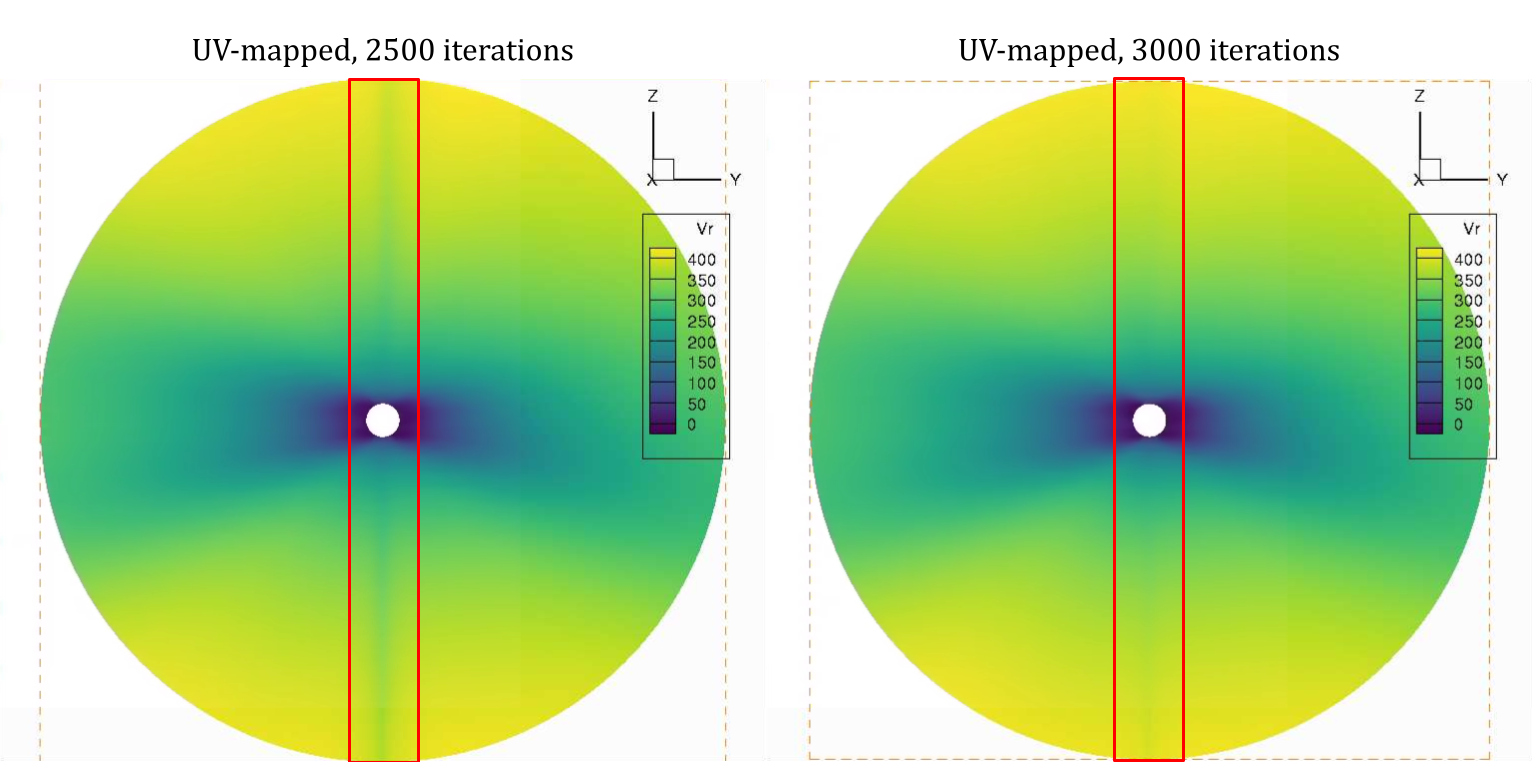}
  \caption{Display of the convergence challenges of a coronal MHD simulation of the 1995 magnetogram when a mesh with highly distorted polar regions is used, in this case introducing spurious outflow during convergence.}
\label{fig:convergenceUV}
\end{figure}

The UV sphere simulation during two different stages of convergence is shown in Figure \ref{fig:convergenceUV}. Despite the fact that there should be very little to almost no outflow or other dynamics around the poles, spurious outflow around the poles (in the red box) is generated at the beginning of the convergence process due to the high local mesh distortion. This then takes a few hundred iterations to be removed from the solution (see the comparison with the solution 500 iterations later, on the right). This issue is not present if the icospheric mesh is used instead, as the distortion due to the knots is not as significant.

\begin{figure}
  \centering
  \includegraphics[scale=0.55]{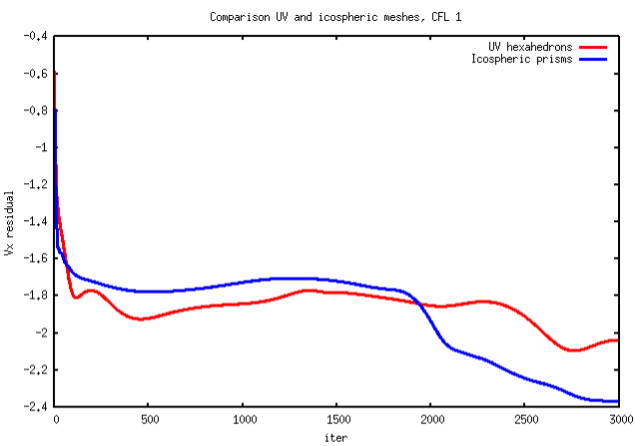}
  \caption{Comparison of a partial convergence history for the 1999 magnetogram in $V_x$ (x component of the velocity) for the UV mesh (\#1) an the fine icospheric mesh (\#2) with a CFL of 1. The beginning of the convergence history is shown to display the effect of the polar regions resulting in the residual oscillations in case of the UV mesh from the iteration 2000 onward.}
\label{fig:CFL1}
\end{figure}

\begin{figure}
  \centering
  \includegraphics[scale=0.37]{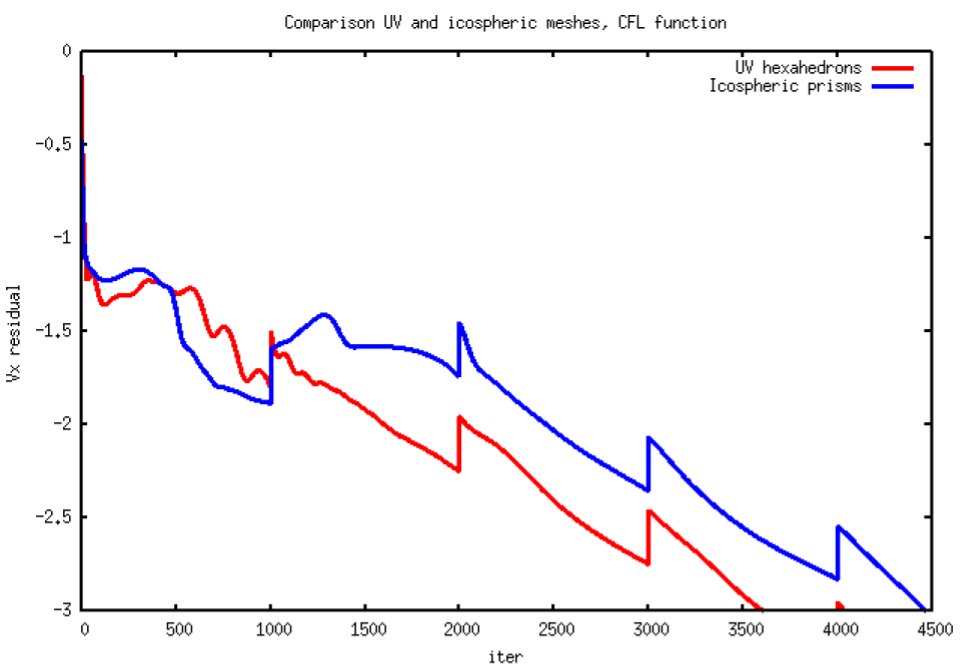}
  \caption{Comparison of the convergence history for the 1999 magnetogram in $V_x$ (x component of the velocity) for the UV mesh (\#1) an the fine icospheric mesh (\#2). The ripples in the convergence curves are caused by the doubling of the CFL every 1000 iterations. The UV mesh reaches the target residual much earlier, mostly due to the fact that a much coarser grid results in a higher numerical dissipation. The oscillations in the UV mesh run are seen to occur for several hundreds of iterations, between iteration 500 to 1500.}
\label{fig:CFLfunction_sameradial}
\end{figure}

This phenomenon can be also identified in the convergence curve of the map from 1999 discussed previously (here the residual is computed from the $V_x$ component), see Figure \ref{fig:CFL1}. While the icospheric mesh (\#2) residual starts decreasing monotonically after around 1500 iterations, oscillations in the residual can be seen in the case of the UV mesh (\#1), precisely because of this polar region problems. For this simulation, a CFL number of 1 for both meshes was used specifically to illustrate this issue and to be able to make a fairer comparison.

The simulation was also ran for longer to achieve proper convergence with a much higher CFL progression, starting from $4$ and then being doubled each 1000 iterations. The convergence results for the $V_x$ component is presented in Figure \ref{fig:CFLfunction_sameradial}. The oscillations due to the polar regions are still apparent in case of the UV mesh between 500 to 1500 iterations, whereas there are no such oscillations present in the icospheric mesh curve.

From this convergence history, intriguingly, it seems as if the UV mesh performed better anyway as it reaches the $V_x$ residual of $< -3$ earlier than its icospheric counterpart, despite the oscillation. Note that here, this residual is defined as $\rm{res}(V_x) = \rm{log}\sqrt{\sum_i\left(V_{x,i}^t - V_{x,i}^{t+1}\right)^2}$ with $i$ being the index for space and $t$ the index for time. Another factor must be considered here as well, however, to explain this phenomenon, and that is the overall mesh size.  

Since the radial distribution is the same and the surface resolution of the icospheric mesh (\#2) much finer, the total number of elements is 3.9M for the icospheric mesh and only 1.7M for the UV mesh (see Table \ref{tab:meshes} and Section \ref{sec:methods}). Coarser mesh can be seen as a source of additional numerical dissipation, which means generally easier convergence. In this case, it is impossible to match the number of cells and the radial discretisation simultaneously between the two grids due to the different topologies. The one level below the current surface discretisation of the icosphere would create a mesh that is 4x coarser, which would not be sufficient to capture the magnetogram features appropriately. 

\begin{figure}
  \centering
  \includegraphics[scale=0.37]{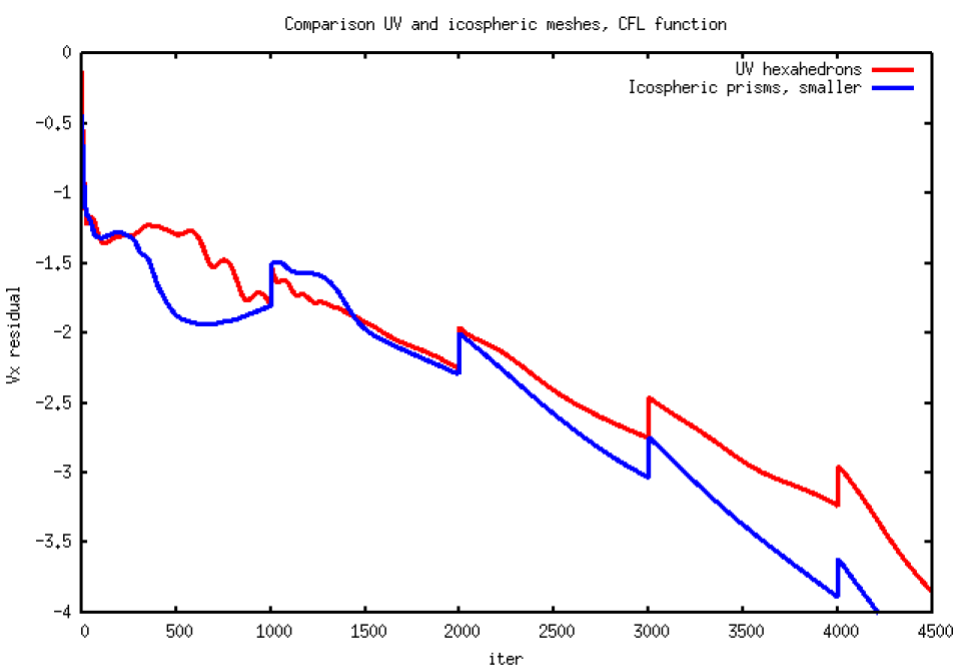}
  \caption{Comparison of the convergence history for the 1999 magnetogram in $V_x$ (x component of the velocity) for the UV mesh (\#1) an the coarser icospheric mesh (\#3). The ripples in the convergence curves are caused by the doubling of the CFL every 1000 iterations. Here, the icospheric mesh reaches the target residual earlier than mesh \#2 from Figure \ref{fig:CFLfunction_sameradial} thanks to a more comparable mesh-associated numerical dissipation.}
\label{fig:CFLfunction_smallerico}
\end{figure}

Thus, a similar, but much smaller icospheric mesh was also used, with the same surface discretisation but with coarser radial discretisation, adding up to roughly 1.33M elements (\#3) and thus being comparable to the UV mesh (\#1) (see Table \ref{tab:meshes} and Section \ref{sec:methods}). Their convergence comparison is shown in Figure \ref{fig:CFLfunction_smallerico}, again for the $V_x$ component. From this plot, it is apparent that the convergence of the icospheric topology is better than that of the UV topology when the level of the mesh-associated numerical dissipation is similar.

For these three grids, the results of the timings are shown in Table \ref{tab:timing}. Using the two icospheric meshes, an equivalent mesh was created through linear interpolation which had the same number of elements as the UV mesh for easier comparison. From this approximation, it is apparent that the icospheric simulation is roughly 20\% faster for the same number of cells. This is also indicated by the number of Krylov sub-iterations (within the Generalised Minimal RESidual algorithm solving the linear discretised system) between subsequent simulation steps that are required, which is much higher for the UV mesh compared to its icospheric counterparts. 

\begin{table}
  \begin{center}
\def~{\hphantom{0}}
\begin{tabular}{lcccc}
Topology                  & Elements & No. elements & $t_\text{CPU}(s)$/ 1000 steps & $\Bar{N}_\text{iters}$/ step \\
UV                        & Hexa     & 1.71M        & 3830                    & 30                          \\
Icospheric                & Prisms   & 3.91M        & 7070                    & 25                          \\
Icospheric                & Prisms   & 1.33M        & 2440                    & 20                          \\
Icospheric (interp.) & Prisms   & 1.71M        & 3120                    & 21                         
\end{tabular}
  \caption{The performance of the different grids with which the coronal MHD 1999 magnetogram-based simulation was computed.}
  \label{tab:timing}
  \end{center}
\end{table}

\section{Discussion}
\label{sec:discussion}

From Section \ref{sec:results}, it is clear that both topologies have their strengths and weaknesses. In Section \ref{sec:introduction}, four different criteria where introduced to aid the mesh trade-off process for the coronal MHD simulations:

\begin{itemize}
    \item The resolution distribution
    \item The resolution adaptability
    \item The convergence performance
    \item The size of the mesh artefacts
\end{itemize}

These will be elaborated on in more detail here after, also considering the results from Section \ref{sec:results}. Subsequently, recommendations will be drawn.

\subsection{Resolution Distribution}

In the context of the coronal MHD simulations for modelling space weather, it is essential that sufficient resolution is available near the equatorial regions. The default UV topology provides the opposite - for a constant angular distribution, which is used in most structured-grid solvers, the highest resolution is actually available in the polar regions and the elements around the equator are the largest. In case of the UV topology, this could be remedied by applying a varying angular distribution to match the cell sizes more closely, or even by concentrating the lines of latitude around the equator to create a positive refinement bias near this region. Such modification could be however time consuming or even impossible to implement for some solvers. Furthermore, it would also have to be considered separately for each magnetogram because especially the coronal simulations of the solar maxima might have also very strong features in the polar regions which could in turn influence the structures near the equator. In addition, even if one managed to define the latitudinal discretisation such that the cells would have roughly the same volume everywhere on the spherical surface, the aspect ratio of the cells would still be very different. Such a mesh cell would have a small extent in the longitudinal direction near the equator compared to its latitudinal extent and vice versa near the poles, biasing the resolution in the domain that way.

From this perspective, the geodesic polyhedron-based topology has an advantage, since even without additional manipulation of the mesh, the surface cell sizes are by default approximately the same everywhere on the surface. Thus, there are no regions with clear refinement advantage.

\subsection{Surface Resolution Adaptability}

Some coronal simulations are based on magnetograms with fairly fine features, resulting in strong streamers and gradients. Hence, in such cases, higher surface refinement might be necessary to capture all this phenomena. From this perspective, the icospheric mesh has a clear disadvantage. Since the surface elements are generated by splitting of the coarser level elements, the number of the elements in the mesh cannot be arbitrary. In the performed simulations, the level-6 division was applied in both icospheric grids (\#2 and \#3). This means that since the level-6 subdivision has 20480 surface elements, the level-7 would have 81920, and the mesh size would hence increase 4 times for the same radial spacing, resulting in over 15M elements. On the other hand, the level-5 mesh would have only 5120 surface cells and so not even a million cells in the full 3D domain, which was found to be unsatisfactory with even the simplest magnetograms. Thus, in this sense, the icospheric mesh is not very adaptable when it comes to the longitudinal and latitudinal resolution. 

Obviously, the mesh could have been adapted such that fewer elements would be used further away from the surface or only some surface elements would be split further, but it is likely that introducing such non-uniformities would only enhance the presence of the spurious fluxes, as was observed in case of the polar regions of the UV mesh where such manipulation was performed.

Note that the radial discretisation is not addressed here since this paper focuses on surface topology. The radial discretisation (and thus radial resolution) is completely independent of the surface grid, and thus not a factor in the trade-off.

\subsection{Convergence Performance}

Due to the highly distorted polar regions, it was shown in Table \ref{tab:timing} that the icospheric mesh has a superior convergence performance when the mesh size is similar. It does not have oscillations in its residual due to spurious outflows near the heavily distorted regions, the iteration steps require fewer sub-iterations and also the CPU time per iteration is shorter. 

\subsection{Mesh Artefacts Size}

Finally, the size of the mesh artefacts should be touched upon. For the applied topologies, the mesh artefacts were only observable in the $\theta$ and $\phi$ components of the velocity field. The case of a rotating dipole showed the significance of these artefacts due to the knots in the icospheric mesh (the regions were the Goldberg polyhedron has pentagons instead of hexagons) in relatively feature-less simulations. The artefacts were mostly contained to lines connecting these knots, but locally creating deviations of up to 20\% in $V_\theta$ and of up to 40\% in $V_\phi$. 

However, when more dynamics was introduced to the simulation by doubling the rotation speed, the relative strength of these artefacts significantly decreased. In the case of a magnetogram-based coronal simulation, the artefacts in $V_\phi$ and $V_\theta$ could be barely seen since it was the strong dynamics of the coronal features that dominated the velocity field. The features were smooth and apart from one line on the example $V_\theta$ isosurface, the icospheric mesh structure could not be recognised in the solution. 

\begin{figure}
  \centering
  \includegraphics[scale=0.25]{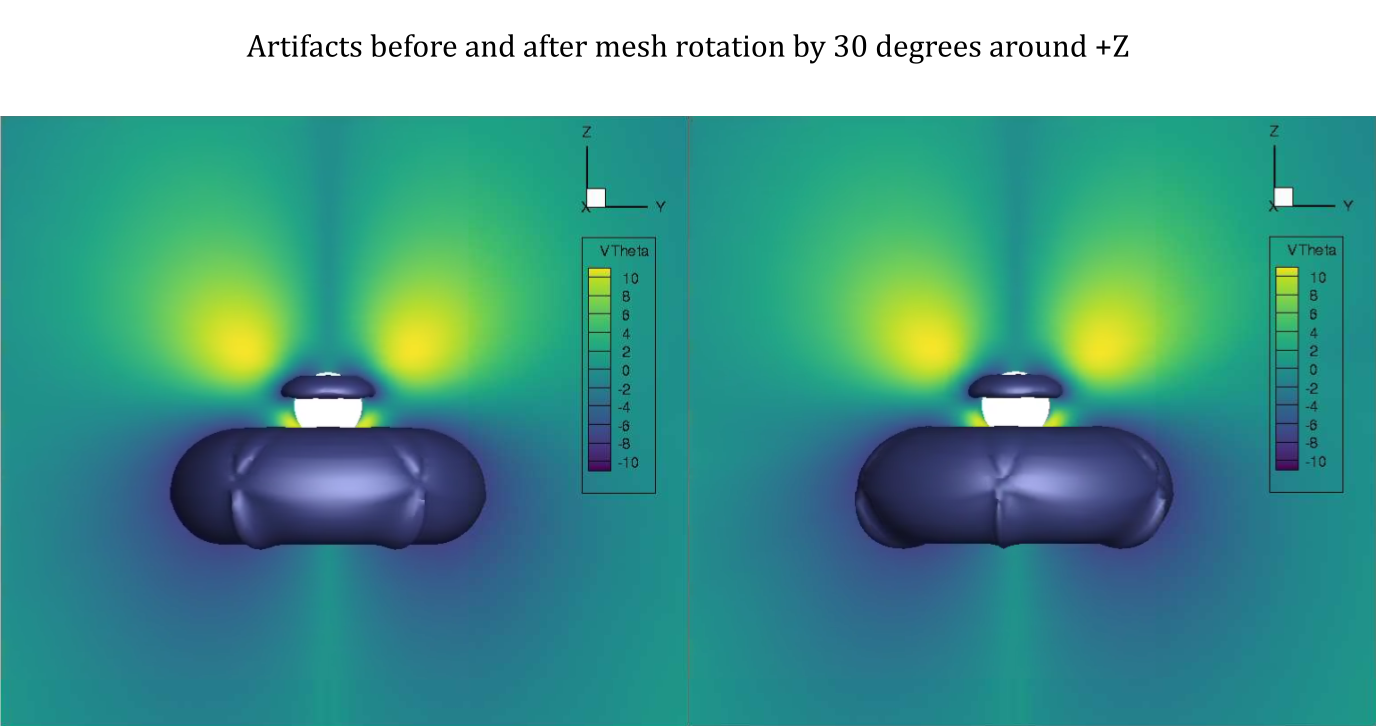}
  \caption{Comparison of the $V_\theta$ artefacts in the solution of the dipole before and after the mesh is rotated by 30 degrees around the Z-axis. The artefacts are now located at different places, allowing the user to identify these features as being indeed mesh-related and analyse the local solution field without the effects of these artefacts.}
\label{fig:Icoshift}
\end{figure}

Should the presence of the knots significantly distort a region of high importance, it is always possible to rotate the mesh such that different regions are distorted and compose the final solution from the two separate simulations made in this way. \textcolor{black}{For the demonstration of this principle, the dipolar solution with the $V_\theta$ artefacts is shown in Figure \ref{fig:Icoshift}, where the mesh was rotated by 30 degrees around the positive Z-axis. It is seen that now the mesh artefacts are at different locations. Thus, it is much easier to identify these features as mesh artefacts and also to see what the solution at these locations would have been had they not been present (in this case symmetric around the Z-axis, as expected). }

The application on the real magnetogram case also illustrated a problem with the artefacts of the UV mesh. For the dipole, there are no features present around the polar regions, thus the effect of the polar distortion was not readily observable in Figure \ref{fig:VthetaDipole} or \ref{fig:VphiDipole}. In case of a magnetogram-based simulation however, these polar-region-associated distortion caused quite significant deformation of the $V_\theta$ structures. While the polar zones are not typically the primary focus of coronal simulations when it comes to space weather applications, especially during the solar maxima these regions might result in strong features affecting the dynamics of the flow also closer to the equator. Thus, the effects of these possible polar artefacts should be evaluated even when only the solution around the equator is considered to be of significance.

\vspace{0.3cm}
\textcolor{black}{All of the conclusions drawn in the four subsections above are shortly summarised in Table \ref{tab:summary}}.

\begin{table}
\begin{tabular}{llll}
                        & Icospheric                                                                                                                   & \hspace{0.2cm}  & UV mapped                                                                                                                                                              \\
Resolution Distribution & \begin{tabular}[c]{@{}l@{}}- uniform everywhere \\ \end{tabular}                                                       &  & \begin{tabular}[c]{@{}l@{}}- by default biased towards the \\ regions of lower interest\end{tabular}                                                                    \\ \\
Resolution Adaptability & \begin{tabular}[c]{@{}l@{}}- very difficult\\ - only in levels\end{tabular}                                                     &  & \begin{tabular}[c]{@{}l@{}}- fairly straightforward \\ - by controlling the angular \\ discretisation\end{tabular}                                                     \\ \\
Convergence Performance & \begin{tabular}[c]{@{}l@{}}- uniform everywhere\\ - without oscillations\end{tabular}                                          &  & \begin{tabular}[c]{@{}l@{}}- difficult in polar regions\\ - creating oscillations\end{tabular}                                                                         \\ \\
Mesh Artefact Size      & \begin{tabular}[c]{@{}l@{}}- exist also in the simplest \\ dipole cases\\ - unnoticeable in flows with \\ strong dynamics\end{tabular} & & \begin{tabular}[c]{@{}l@{}}- non-observable in cases without \\  polar outflow\\ - significant when polar \\ outflow present, even with strong \\ dynamics\end{tabular}
\end{tabular}
  \caption{The summary of the performance of the two mesh topologies according to the four criteria selected for investigation.}
  \label{tab:summary}
\end{table}

\subsection{Recommendations}

The major challenge associated with using the icosphere was related to the relatively major spurious flows in otherwise feature-less simulations (such as the slow rotating dipole), with the artefacts mostly located around the knot lines. While it is simple to identify these artefacts in the solution of a dipole, telling the spurious and physical fluxes apart could be much more challenging for a magnetogram-based simulation with strong features (even though it was observed that with these strong features, the relative strength of the mesh artefacts was not large).  

In this case, a good solution would be to run the case twice for verification purposes, where in the second run, the mesh would be rotated in the polar direction by a few degrees. This way, the locations of the knots and the knot lines would move and affect a different part of the solution domain, as shown in Figure \ref{fig:Icoshift}. Thus, from the comparison between these two results, the effects of the artefacts could be isolated and, if needed, removed. 

The major drawbacks of using the UV mesh were found to be the unfavourable resolution distribution (from the perspective of space weather applications) and the heavily distorted polar regions, where the latter significantly compromised the accuracy of the results near the poles and also affected the convergence of the simulation.

The unfavourable resolution distribution could be mitigated by re-mapping of the UV sphere, where constant angular spacing would be abolished and the lines of the latitude would be concentrated closer together near the equator. While this would make the mesh generation and handling more complicated, it could allow for even coarser meshes and more efficient computations. However, even then there would still be a resolution bias, since the aspect ratio of the resulting cells would vary greatly across the domain.

A possibility to improve the UV mesh distortion is to further experiment with methods to deal with the degenerate elements near the poles. For example, the domain can be split into two segments with additional boundary conditions at the poles (see e.g. \citet{Perri2020}), such that the degenerated regions are omitted. Whether such a solution leads to physical results is, however, debatable. 

In summary, since the polar region distortion caused highly observable spurious flows also in the magnetogram-based simulations and since the UV-mapped mesh was observed to have worse convergence, based on our results, it is generally recommended to use the icospheric mesh (geodesic polyhedron-based), possibly with a simulation re-run after the mesh is angularly shifted in the polar direction to isolate the spurious fluxes.

\section{Conclusions}
\label{sec:conclusions}

The purpose of this work was to discuss the different mesh topologies applicable for coronal MHD simulations and outline their benefits and disadvantages. The coronal MHD simulations had a spherical domain of $21 R_\text{Sun}$ with a spherical $R_\text{Sun}$ cut-out in the middle, representing the Solar surface. 

Two basic surface mesh topologies were investigated; a topology based on the UV mapping (here referred to as a UV sphere) and a topology based on the geodesic polyhedron (here referred to as an icospheric mesh). The former has a constant angular spacing between the quadrilateral elements, while the latter has equally sized triangular elements on the surface. The UV sphere has distorted regions near the poles, where the originally quadrilateral elements degenerate, whereas the icospheric mesh has what we refer to as knots on the surface in the regions where its dual construction, the Goldberg polyhedron, contains pentagons instead of hexagons.

The configurations were discussed from a variety of perspectives, such as their computational performance, adaptability, resolution distribution and the size of the mesh artefacts. The mesh artefacts were caused due to spurious numerical fluxes in simulations close to the hydrostatic equilibrium (with a state-of-the-art FV scheme that was not strictly well-balanced) and in all cases, they were only observable in the $\theta$ and $\phi$ velocity components. The COOLFluiD platform with a steady-state implicit scheme was used for the ideal MHD simulations.

Firstly, a simple rotating magnetic dipole was computed, showing relatively significant spurious fluxes in case of the icospheric mesh, mainly on the lines connecting the nonuniform knot regions. Afterwards, the same rotating dipole was simulated with double the rotation speed, which revealed that while the spurious fluxes were still present, their relative magnitude was much smaller compared to the strength of the actual physical features. This demonstrated that these mesh artefacts become less significant if there are stronger structures in the solution. From this,  it is inferred that these artefacts might not be so problematic for simulations with strong dynamics, such as those based on magnetograms instead of simple dipole configurations. 

Therefore, also magnetogram-based simulations were performed, using magnetogram data from 1999 (solar maximum) and 1995 (solar minimum). The results from 1999 were investigated to evaluate the significance of the mesh effects. The icospheric mesh artefacts were hardly detectable. On the other hand, since the coronal features here extended well into the polar regions where the UV mesh has a large non-uniformity, it was the UV mesh-based solution that suffered from strong artefacts in these zones. 

Afterwards, the convergence was also assessed. Here, monitoring the convergence progress of the 1995 simulation revealed that the UV mesh suffers from worse convergence due to the highly distorted polar regions. This showcased itself in the convergence curve of the 1999 map in the form of oscillations in the residual. Timing of the simulation using a variety of grids and interpolation revealed that the icospheric mesh converges faster with fewer sub-iterations needed between the iteration steps. 

In the discussion, it has been elaborated on that the topology which is more appropriate for a coronal MHD simulation highly depends on the simulation features. For relatively feature-less simulations close to a hydrostatic equilibrium, the icospheric mesh might introduce too many artefacts in the solution if the scheme is not well balanced. The UV mesh has much smoother features near the equator and might be hence more appropriate if the polar regions are not of interest.

On the other hand, in magnetogram-based simulations where stronger features are present in the domain including in the polar regions, the icospheric mesh provides the benefit of almost equally-sized cells everywhere on the surface and no regions with of high refinement or distortion bias. It also converges faster than the UV mesh for the same number of elements.

Some recommendations were also formulated on how to further improve the mesh performance. For the UV mapped mesh, the lines of latitude could be concentrated closer towards the equator to better resolve the actual region of interest and thus run more efficiently. The distortion in the polar zones could be reduced by further experimenting with the transformation of the degenerate prism elements, or by splitting the domain into two segments with additional boundary conditions. For the icospheric mesh, if the possible effects of the mesh artefacts on the solution cannot be estimated a priori, two separate simulations could be run, with the mesh being rotated by a few degrees angularly such that the knot lines affect a different portion of the domain. Comparison of such two solutions would allow for isolation of the mesh effects and provide a reliable solution.

\section*{Acknowledgements}
The corresponding author thanks the referees for their constructive feedback. 

\section*{Funding}
This work has been granted by the AFOSR basic research initiative project FA9550-18-1-0093. This project has also received funding from the European Union’s Horizon 2020 research and innovation programme under grant agreement No 870405 (EUHFORIA 2.0) and the ESA project "Heliospheric modelling techniques“ (Contract No. 4000133080/20/NL/CRS). F.Z. is supported by a postdoctoral mandate from KU Leuven Internal Funds  (PDMT1/21/028).
These results were also obtained in the framework of the projects C14/19/089  (C1 project Internal Funds KU Leuven), G.0D07.19N  (FWO-Vlaanderen), SIDC Data Exploitation (ESA Prodex-12), and Belspo projects BR/165/A2/CCSOM and B2/191/P1/SWiM. The resources and services used in this work were provided by the VSC (Flemish Supercomputer Centre),
funded by the Research Foundation - Flanders (FWO) and the Flemish Government.

\section*{Declaration of interests}
The authors report no conflict of interest.

\section*{Data availability statement}
The data supporting the results of this study can be made available from the corresponding author upon request.

\clearpage
%\section*{Authors ORCID}
%M. Brchnelova https://orcid.org/0000-0003-0874-2669, F. Zhang %https://orcid.org/0000-0002-9425-994X, P. Leitner %https://orcid.org/0000-0003-3792-0452, B. Perri %https://orcig.org/0000-0002-2137-2896, A. Lani %https://orcig.org/0000-0003-4017-215X, S. Poedts %https://orcig.org/0000-0002-1743-0651

%All papers included in the References section must be cited in the article, and vice versa. Citations should be included as, for example ``It has been shown \citep{Rogallo81} that...'' (using the {\verb}\citep} command, part of the natbib package) ``recent work by \citet{Dennis85}...'' (using {\verb}\citet}).
%The natbib package can be used to generate citation variations, as shown below.
%\begin{itemize}
%\item \verb#\citet[pp. 2-4]{Galtier00}#:
%\citet[pp. 479-480]{Galtier00} 
%\item \verb#\citep[p. 6]{Worster92}#:
%\citep[p. 6]{Worster92}
%\item \verb#\citep[see][]{Koch83, Lee71, Linton92}#:
%\citep[see][]{Koch83, Lee71, Linton92}
%\item \verb#\citep[see][p. 18]{Martin80}#:
%\citep[see][p. 18]{Martin80}
%\item \verb#\citep{Brownell04,Brownell07,Ursell50,Wijngaarden68,M%iller91}#:
%\citep{Brownell04,Brownell07,Ursell50,Wijngaarden68,Miller91}
%\end{itemize}

%\appendix

% susie put cite commands here, don't bother with citet etc just yet.

\bibliographystyle{jpp}

\bibliography{jpp-instructions}

\end{document}